\documentstyle[12pt,aaspp4]{article}
\def\apj #1 #2 #3 {#1, ApJ, {\bf #2}, #3}
\def\apjl #1 #2 #3 {#1, ApJ, {\bf #2}, L#3}
\def\apjs #1 #2 #3 {#1, ApJS, {\bf #2}, #3}
\def\aap  #1 #2 #3 {#1, A\&A, {\bf #2}, #3}
\def\mnras #1 #2 #3 {#1, MNRAS, {\bf #2}, #3}
\def\pra #1 #2 #3 {#1, Phys.~Rev.~A., {\bf #2}, #3}
\def\prb #1 #2 #3 {#1, Phys.~Rev.~B., {\bf #2}, #3}
\def\prc #1 #2 #3 {#1, Phys.~Rev.~C., {\bf #2}, #3}
\def\prd #1 #2 #3 {#1, Phys.~Rev.~D., {\bf #2}, #3}
\def\pre #1 #2 #3 {#1, Phys.~Rev.~E., {\bf #2}, #3}
\def\prl #1 #2 #3 {#1, Phys.~Rev.~Lett., {\bf #2}, #3}
\def\plb #1 #2 #3 {#1, Phys.~Lett.~B., {\bf #2}, #3}
\def\science #1 #2 #3 {#1, Science., {\bf #2}, #3}
\def\nature #1 #2 #3 {#1, Nature., {\bf #2}, #3}
\def\nphysa #1 #2 #3 {#1, Nucl.~Phys.~A., {\bf #2}, #3}
\def\nphysb #1 #2 #3 {#1, Nucl.~Phys.~B., {\bf #2}, #3}
\def\nphysbs #1 #2 #3 {#1, Nucl.~Phys.~B.~Suppl., {\bf #2}, #3}
\def\eprint{Los Alamos e-Print Archive., }
\def\h#1{\hbox{${}^{#1}$H}}
\def\he#1{\hbox{${}^{#1}$He}}
\def\li#1{\hbox{${}^{#1}$Li}}
\def\be#1{\hbox{${}^{#1}$Be}}
\def\b#1#2{\hbox{${}^{#1#2}$B}}
\def\dovh{\hbox{D/H}}

\def\yp{\hbox{\hbox{$Y_p$}}}

\def\omegab{\hbox{${\sl\Omega}_{\rm b}$}}
\def\h502{\hbox{$ h^{2}_{50}$}}
\def\xinue{\hbox{$\xi_{\nu_{e}}$}}
\def\xinum{\hbox{$\xi_{\nu_{\mu}}$}}
\def\xinut{\hbox{$\xi_{\nu_{\tau}}$}}
\def\xinumt{\hbox{$\xi_{\nu_{\mu, \tau}}$}}
\def\nue{\hbox{$\nu_{e}$}}
\def\num{\hbox{$\nu_{\mu}$}}
\def\nut{\hbox{$\nu_{\tau}$}}
\def\ev{\mbox{~eV}}
\def\mev{\mbox{~MeV}}

\def\la{\mathrel{\mathpalette\fun <}}
\def\ga{\mathrel{\mathpalette\fun >}}
\def\fun#1#2{\lower3.6pt\vbox{\baselineskip0pt\lineskip.9pt
  \ialign{$\mathsurround=0pt#1\hfil##\hfil$\crcr#2\crcr\sim\crcr}}}
\begin{document}
\bigskip
\bigskip
\rightline{NAOJ-Th/}
\rightline{astro-ph/0005446}
\title{
Neutrino Degeneracy and Decoupling: New Limits from
Primordial Nucleosynthesis and  the Cosmic Microwave Background
}
\author{
M. Orito\altaffilmark{1}, T. Kajino\altaffilmark{1,2,3},
G. J. Mathews\altaffilmark{1,4},
and  R. N. Boyd\altaffilmark{5}
}
\altaffiltext{1}{
National Astronomical Observatory, 2-21-1, Osawa, Mitaka, Tokyo 181-8588, Japan
}
\altaffiltext{2}{
Graduate University for Advanced Studies, Department
of Astronomical Science, 2-21-1, Osawa, Mitaka, Tokyo 181-8588, Japan
}
\altaffiltext{3}{
University of Tokyo, Department of Astronomy, 7-3-1
Hongo, Bunkyo-ku, Tokyo 113-0033, Japan
}
\altaffiltext{4}{
Center for Astrophysics, and 
Department of Physics, University of Notre Dame, Notre Dame,
IN 46556
}
\altaffiltext{5}{
Department of Physics, Department of Astronomy, Ohio
State University, Columbus, OH 43210
}
\begin{abstract}
We reanalyze the cosmological constraints on the existence of
a net universal  lepton asymmetry and neutrino degeneracy.  
We show that neutrinos can begin to decouple at higher temperatures
than previous estimates due to several corrections which diminish the
neutrino reaction rate.
For sufficiently large degeneracy, neutrino
decoupling can occur before various particles annihilate
and even before the QCD phase transition.
These decoupled neutrinos are therefore not heated as the particle
degrees of freedom change.
The resultant ratio of the relic neutrino-to-photon temperatures
after $e^\pm$ annihilation can then be significantly reduced
by more than a factor of two from that of the standard nondegenerate 
ratio.  This changes the expansion rate and subsequent primordial
nucleosynthesis, photon decoupling, and structure formation.  
In particular we analyze physically plausible
lepton-asymmetric models with 
large $\nu_\mu$ and $\nu_\tau$  degeneracies
together with a moderate $\nu_e$ degeneracy.  We show
that the nucleosynthesis by itself permits very 
large neutrino degeneracies  $0 \le \xinum$, $\xinut \le 40$,
$0 \le \xinue \le 1.4$ together with large 
baryon densities $0.1 \le \Omega_b \h502 \le 1$ 
as long as some destruction of primordial lithium has occurred.
We also show that structure formation
and the power spectrum of the cosmic microwave background
allows for the possibility of an
$\Omega = 1$, $\Omega_\Lambda = 0.4$, cosmological model 
for which there is both significant lepton asymmetry
($\vert \xinum \vert = \vert  \xinut \vert   \approx 11$) and  a relatively
large baryon density ($\Omega_b \h502 \approx 0.2$). 
Our best-fit neutrino-degenerate, high-baryon-content 
models are mainly distinguished by a suppression of the
second peak in the microwave background power spectrum.
This is consistent with the recent high resolution
data from BOOMERANG and MAXIMA-1.
\end{abstract}
\keywords{
cosmology: lepton asymmetry, $\omegab$ -  cosmic microwave background}
\section{INTRODUCTION}
%
%

The universe appears to be charge neutral to very
high precision (\cite{lyttleton59}). Hence, any 
universal net lepton number beyond the
net baryon number must reside entirely within the neutrino sector.  Since
the present relic neutrino number asymmetry is not directly observable
there is no firm experimental basis for postulating that the lepton number
for each species is zero. It is therefore possible for the individual lepton
numbers $L_l$ of the universe to be large compared to the 
baryon number of the
universe, $B$, while the net total lepton number is small $(L \sim B)$.
Furthermore, it has been suggested that even if the baryon number
asymmetry is small, the total lepton number could be large in the context
of $SU(5)$ and $SO(10)$ Grand Unified Theories (GUT's)
(\cite{harvey81}; \cite{fry82}; \cite{dolgov91}; \cite{dolgov92}).
It has also been proposed
recently (\cite{casas99}) that models based upon the Affleck-Dine scenario
of baryogenesis (\cite{affleck85}) might 
generate naturally lepton
number asymmetry which is seven to ten orders of magnitude 
larger than the baryon number
asymmetry. Neutrinos with large lepton asymmetry and masses
$\sim 0.07 \ev$ might even explain the 
existence of cosmic rays  with energies in excess of
the Greisen-Zatsepin-Kuzmin cutoff (\cite{greisen66}; \cite{gelmini99}).
Degenerate, massive (2.4 eV) neutrinos may even be required (\cite{larsen95})
to provide a good fit to the power spectrum of large scale
structure in mixed dark matter models. It is, therefore, equally important
for both particle physics and cosmology to carefully scrutinize
the limits which cosmology places on the allowed range of both the lepton and
baryon asymmetries.

The consequences 
of a large lepton asymmetry and associated neutrino degeneracy
for big bang nucleosynthesis (BBN)
have been considered in many papers.  Models with degenerate 
$\nue$ (\cite{wagoner67}; \cite{terasawa85}; 
\cite{scherrer83}; 
\cite{terasawa88};
\cite{kajino98a}), 
both $\nue$ and $\num$ (\cite{yahil76};
\cite{beaudet76}; \cite{beaudet77}), and for three degenerate
neutrino species
(\cite{david80}; \cite{olive91}; \cite{kang92}; \cite{starkman92};
\cite{kajino98b}; \cite{kim95}; \cite{kim98}) have been analyzed. 
The effects of the
degeneracy of electron type neutrinos on inhomogeneous BBN models
were also considered in  Kajino \& Orito (1998a, 1998b).
Constraints on lepton asymmetry also arise from the requirement
that sufficient structure develop by the present time (\cite{kang92})
and from the power spectrum of fluctuations in 
galaxies (\cite{larsen95}) and the cosmic microwave
background temperature (\cite{kinney}; \cite{lesgourgues}; 
\cite{hannestad}).

The present work differs from those listed above
primarily in that we carefully reexamine the neutrino decoupling
temperature.  
Although neutrino degeneracy has been treated in considerable detail
previously (cf. \cite{kang92}) we reexamine several issues
which were not treated in that paper or elsewhere.  We show that there are
corrections to the neutrino reaction rates which
cause the neutrino decoupling temperature to be significantly
higher than previous estimates.  These corrections include
a proper accounting of effects of degeneracy on the
abundances of the weakly-reacting species present.
We also consider corrections 
in the the electron and photon effective masses and 
the relative velocities of interacting species.  
In addition we make a more careful continuous 
treatment of the various particle annihilation epochs.
These  corrections
are important since it becomes easier for neutrinos to decouple before
the annihilation of various particles and even  before the QCD transition.
The fact that neutrinos may decouple when there were many particle
degrees of freedom causes the relic neutrino temperature to be much lower
than that of the standard nondegenerate big bang.
This allows for new 
solutions for BBN in which significant baryon density is possible
without violating the various abundance constraints.
We find that for decoupling temperatures just above 
the QCD epoch it is also possible to find models in which the 
structure constraint and
even the CMB power spectrum constraint are marginally satisfied.

In this paper we explore a physically plausible range of baryon and lepton
asymmetric cosmological models in which 
the electron neutrino is only slightly degenerate
while the muon and tau neutrinos have nearly equal (but opposite-sign)
significant chemical potentials.  We use these models to deduce new cosmological  
constraints on the baryon and lepton content of the universe.  
We emphasize that
previous studies of BBN and the CMB temperature fluctuations
with highly degenerate neutrinos have not
exhaustively scrutinized the possibility of
important effects from the lower implied 
neutrino temperature  when neutrinos decouple before various particle
annihilations and/or the QCD epoch. 

In what follows we first discuss here how the observable BBN 
yields in a neutrino-degenerate
universe impose bounds on the baryon and lepton asymmetries
which still allow a large neutrino degeneracy and
baryon density (even $\omegab = 1$). We show that there exists a 
one parameter family of allowed neutrino-degenerate models
which satisfy all of the constraints from BBN. 
We next examine other cosmological 
non-nucleosynthesis constraints, i.e.
 the cosmic microwave 
background (CMB) fluctuations and the time scale
for  the development of structure.
We show that these constraints can be marginally satisfied
for a limited range of highly degenerate
models from the  fact the relic neutrino
temperature is much lower than that of the standard nondegenerate big bang.
We also discuss how a determination of the neutrino degeneracy 
parameters could constrain the neutrino mass spectrum  
from the implied neutrino contribution $\Omega_\nu$ to the closure density.
\section{LEPTON ASYMMETRY \& NEUTRINO DECOUPLING}
In this section, we review the basic relations which define the
magnitude of neutrino degeneracy and summarize
the cosmological implications. 
Radiation and relativistic particles dominate 
the evolution of the early hot big bang.
In particular,
relativistic neutrinos with masses less than the
neutrino decoupling temperature, $m_\nu \la {\cal O}(T_D) \sim
\mev$, contributed an energy density greater than that due to photons
and charged leptons.  Therefore, a small modification of neutrino properties
can significantly change the expansion rate of the universe. 
The energy density of
massive degenerate neutrinos and antineutrinos for each species
is  described by the usual 
Fermi-Dirac distribution functions $f_\nu$ and $f_{\bar\nu}$, 
\begin{eqnarray}
	\lefteqn{\rho_\nu + \rho_{\bar{\nu}} =}
	\nonumber \\
	&&\frac{1}{2\pi^2}\int_0^\infty \!\! \! dp ~p^2 \sqrt{p^2 + m_\nu^2}
                              (f_\nu(p)+ f_{\bar{\nu}}(p)).
	\label{eq:rhonu}
\end{eqnarray}
where, $p$ denotes the magnitude of the 3-momentum, 
and $m_\nu$ is the neutrino mass.
Here and throughout the paper we use natural units( $\hbar=c=k_B=1$).
The distribution functions are
\begin{equation}
	\renewcommand{\arraystretch}{2.0}
	\begin{array}{rcl}
		\displaystyle
		f_\nu(p)=\frac{1}{\exp \left(p/T_\nu -
		\xi_\nu\right)+1}~, \\
		\displaystyle
		f_{\bar{\nu}}(p)=\frac{1}{\exp
		\left(p/T_\nu + \xi_\nu\right)+1}~,
	\end{array}
	\label{eq:FermiDirac}
\end{equation}
where the degeneracy parameter $\xi_\nu$ is defined in term
of the neutrino chemical potential, $\mu_\nu$, 
as $\xi_\nu \equiv \mu_\nu/T_\nu$.  It will have a nonzero
value if a lepton asymmetry exists. 
In the limit of massless neutrinos, the energy density in neutrinos
becomes
\begin{equation}
\rho_{\nu}  + \rho_{\bar{\nu}} = {7 \over 8} {\pi^2 \over 15} \sum_i T_{\nu_i}^4
 \biggl[1 + {15 \over 7} \biggl(
{\xi_{\nu_i} \over \pi}\biggr)^4 + {30 \over 7} \biggl(
{\xi_{\nu_i} \over \pi}\biggr)^2\biggr]~~,
\label{dense}
\end{equation}
from which it is clear that a neutrino degeneracy in any species
tends to increase the energy density.

The net lepton asymmetry $L$ of the universe can
be expressed 
to high accuracy as
\begin{equation}
	\renewcommand{\arraystretch}{2.0}
	\begin{array}{rcl}
		\displaystyle L = \sum_{l=e,\mu,\tau}L_l, \\
		\displaystyle L_l = \frac{n_{\nu_l} -
n_{\bar{\nu}_l}}{n_\gamma}.
		\label{eq:lepasy}
	\end{array}
\end{equation}
This is analogous to the
baryon-to-photon ratio
$ \eta \equiv (n_B - n_{\bar{B}})/n_\gamma$. Here,
$n_{\nu_l}$
$(n_{\bar{\nu}_l})$ 
are the number densities for  each neutrino (anti-neutrino) species, 
$n_\gamma$ is the photon number density, and $n_B$\,($n_{\bar{B}}$)
is the (anti)\,baryon number density.
Once the
temperature drops sufficiently below the muon rest  mass, say  
$T \la 10 \mev$,
all charged leptons except for electrons and positrons
will have decayed away.  Overall charge neutrality 
then demands that the difference between the  number densities
of electrons and positrons equal the proton number density.
Hence,  any electron degeneracy is 
negligibly small and any significant lepton asymmetry must reside
in the neutrino sector. 
After the epoch of $e^{\pm}$ annihilation, the magnitudes 
of the lepton and baryon asymmetries
 are conserved.  They are equal to the present value in
the absence of any
subsequent baryon and/or lepton number-violating processes.

Elastic scattering reactions, such as
$\nu_l\,(\bar{\nu}_l) + l^{\pm} \leftrightarrow \nu_l\,(\bar{\nu}_l) +
l^{\pm}$, keep the neutrinos in kinetic equilibrium.  Annihilation
and creation processes which can change
their number density, like $\nu_l + \bar{\nu}_l \leftrightarrow
l + \bar{l}$, $\nu_l + l' \leftrightarrow \nu_{l'} + l$, etc,
maintain the neutrinos in chemical equilibrium.
When the rates for these weak interactions 
become slower than the Hubble expansion rate,
neutrinos decouple and begin a "free expansion''.  Their  number densities 
continue to diminish as $1/R^3$
and their momenta red-shift by a factor $1/R$, where $R$ is the cosmic
scale factor. However, this decoupling has no effect on
the shape of the distribution functions.  Relativistic neutrinos 
and antineutrinos continue to be described by the Fermi-Dirac
 distributions of Eq.~(\ref{eq:FermiDirac}).
Since the individual lepton number is believed to be conserved, the
degeneracy parameters $\xi_{\nu_l}$ remain constant after decoupling.

\subsection{Neutrino Decoupling Temperature}
When one estimates the
present density of relic neutrinos one must consider
the effect of the changing number of degrees of freedom
for the remaining interacting particles.
For example, once the neutrinos are totally decoupled,
they are not heated by subsequent  pair annihilations. Hence, their
temperature $T_{\nu}$ is lower than the temperature  $T_\gamma$ 
of photons (and any other electromagnetically
interacting particles) by a factor $y_{\nu} = T_{\nu}/T_{\gamma}$.
In the standard non-degenerate cosmology,
with three flavors of
massless, non-degenerate neutrinos which decouple
just before the $e^{\pm}$ pair annihilation epoch,
the present ratio of the neutrino to photon
temperatures is given by  $y_{\nu} = (4/11)^{1/3}$.

Neutrinos and anti-neutrinos drop out of thermal equilibrium with the
background thermal plasma  when
the weak reaction rate per particle, $\Gamma$, 
becomes slower than the expansion rate, $H$.
Since the decoupling occurs quickly, it is a good
approximation to estimate a characteristic
decoupling temperature $T_D$, 
at which the slowest weak-equilibrating reaction per particle $\Gamma$
becomes slower than the expansion rate.  

The expansion rate is simply given by the Friedmann equation
\begin{equation}
H = \sqrt{(8/3) \pi G \sum_i \rho_i}~~,
\label{friedmann}
\end{equation}
where the sum is over all species contributing mass energy (mostly 
relativistic particles).
Obviously, the universal expansion in neutrino-degenerate models
is more rapid because of the higher neutrino
mass-energy density (e.g.~Equation \ref{dense}).
This causes the  decoupling to occur
sooner and at a higher temperature than
in the non-degenerate case. 

Our estimates of the
weak decoupling temperature as a function of neutrino degeneracy are
shown in Figure 1 and compared with the previous 
estimates Kang \& Steigman (1992).
Our decoupling temperatures are
significantly higher than those estimated in Kang \& Steigman (1992) for 
reasons which we now explain.
 
Significant neutrino degeneracy will cause
the weak reaction rate per particle, $\Gamma$, to be slower
than the nondegenerate case because the neutrino final states are occupied 
(\cite{kang92}; \cite{freese83}).  More importantly, the existence of a
lepton asymmetry represented by 
a finite positive neutrino chemical potential will cause the number density
of antineutrinos to be significantly less than that of neutrinos.
Hence, the annihilation rate per neutrino $\Gamma_\nu  = 
n_{\bar \nu} \langle \sigma_{\nu \bar \nu} \rangle$ can be significantly
less than the reaction rate per antineutrino, electron or positron.

In Kang \& Steigman (1992) the weak reaction rate was estimated from
the reaction rate per electron annihilation,
 $e^{+} + e^{-} \rightarrow \nu_i + \bar{\nu_i}$.  We will call this
$\Gamma_e$.  One could however also consider the reaction rate per
positron, $\Gamma_{e^+}$, the reaction rate per neutrino
species $i$, $\Gamma_{\nu_i}$,
or the reaction rate per antineutrino, $\Gamma_{\bar \nu_i}$.
In the limit of high temperature and  no neutrino degeneracy, 
all of these reaction rates are identical.
However, when  for example 
the neutrinos have a large positive chemical potential,
the neutrino annihilation rate $\Gamma_{\nu_i} = 
n_{\bar \nu_i} \langle \sigma v\rangle$ 
becomes the rate determining reaction.  This is 
because of the scarcity of anti-neutrinos with which to interact.
It is then this reaction rate and not $\Gamma_e$
which should be used to determine when the weak reactions are no longer able to
maintain chemical equilibrium.  

In Kang \& Steigman (1992) expressions were derived for
the phase space integrals for $\langle \sigma v\rangle$ 
which reduce to
\begin{equation}
\langle \sigma v\rangle = {8 \bigl(a^2 + b^2\bigr) \over
27 \pi \zeta^2(3)} I(\xi) G_F T^2~~,
\end{equation}
where $I(\xi)$ is a phase space integration factor given in that paper,
$\zeta$ is the Riemann zeta function, and
$G_F$ is the Fermi coupling constant.  The quantities $a$ and $b$
relate to the Weinberg angle, $a = 1/2 + 2 \sin^2{\theta_W}$ for 
electron neutrinos, and $a = 1/2 - 2 \sin^2{\theta_W}$ for muon and
tau neutrinos, while $b = 1/2$.

Using this, we derive the following expression for the 
decoupling temperature corresponding to each neutrino species, $i$,
\begin{equation}
T_D(\xi_i) \approx \biggl[{4.38 \over {a^2  + b^2}} \biggr]^{1/3}
 \biggl[ {\sqrt{g'_{eff}(\xi)} \over I(\xi)}\biggr]^{1/3}
\times \biggl( {n_{\bar i} \over
n_{e}} \biggr)^{1/3}~{\rm MeV}~~~,
\label{tdcorr}
\end{equation}
where $\bar i$ denotes the matter (or antimatter) counter part for
the species under consideration.  The formula
A.8 of Kang \& Steigman (1992)  omits the latter density ratio factor which is
only unity in the limit of no degeneracy.  Thus,   
their $\Gamma$ used to deduce the decoupling temperature was 
too large.  We have checked the above result against a full eight-dimensional
integration of the Boltzmann equation and find our results to be correct.

There are also two other differences between our results and 
those  of Kang \& Steigman (1992):
 The first is that we have included
finite temperature  corrections to the mass of the electron and photon
(\cite{fornengo97}). 
The second is that we have calculated the average neutrino annihilation
rate in the cosmic comoving frame.  In this frame
the M{\o}ller velocity is used instead of
the relative velocity for the integration of the collision term
in the Boltzmann equations (\cite{gondolo91}; \cite{enqvist92}).
These corrections also slightly increase the
neutrino decoupling temperatures.
Therefore even in the nondegenerate case we find 
slightly higher decoupling temperatures than those of Kang \& Steigman (1992):
\begin{eqnarray}
	T_D(\xi_\nu = 0) \simeq 2.33 \mev \quad\rm{for} ~\nu_e~, \nonumber \\
	T_D(\xi_\nu = 0) \simeq 3.90 \mev \quad\rm{for} ~\nu_{\mu, \tau}.
\end{eqnarray}
 These values are in good agreement with those of Enqvist et al. (1992).
If we remove the above corrections, we recover the decoupling 
temperatures of (Kang \& Steigman 1992), i.e. 
$T_D(\xi_\nu = 0) \simeq 1.98$ MeV for $\nu_e~$, and
$T_D(\xi_\nu = 0) \simeq 3.30$ MeV for $\nu_{\mu, \tau}$.

%

\subsection{Relic Neutrino Temperature}
\begin{figure}
\plotone{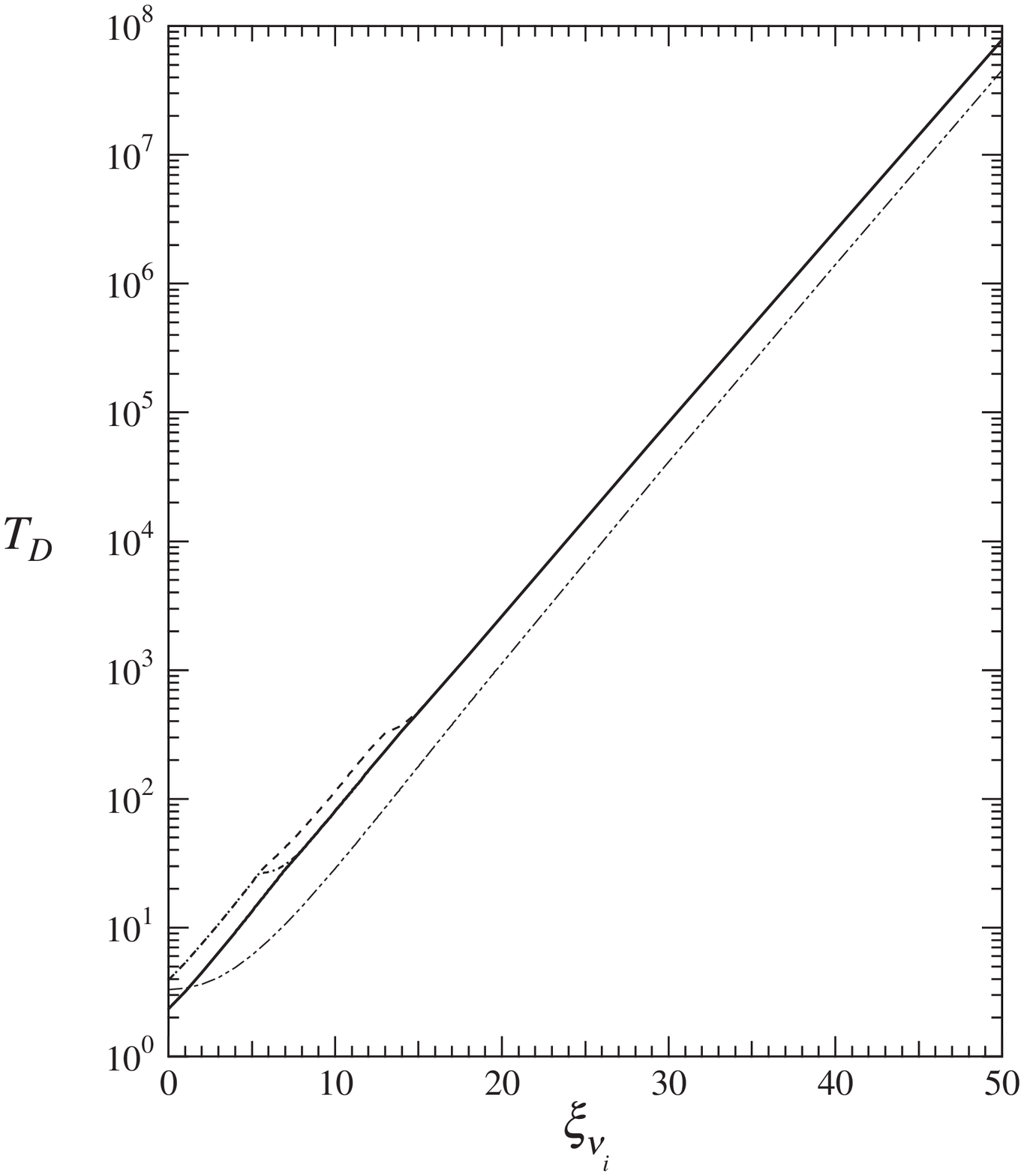}
\caption{Decoupling temperature $T_D$ (in MeV)  for 
$\nu_e$ (solid line), $\nu_\mu$ (dot-dashed line), 
and $\nu_\tau$ (dashed line)    
as a function
of degeneracy parameter $\xi_{\nu_i}$, compared with the previous 
estimate (two-dot dashed line) of Kang \& Steigman (1992)}
\protect\label{fig:1}
\end{figure}

\begin{figure}
\plotone{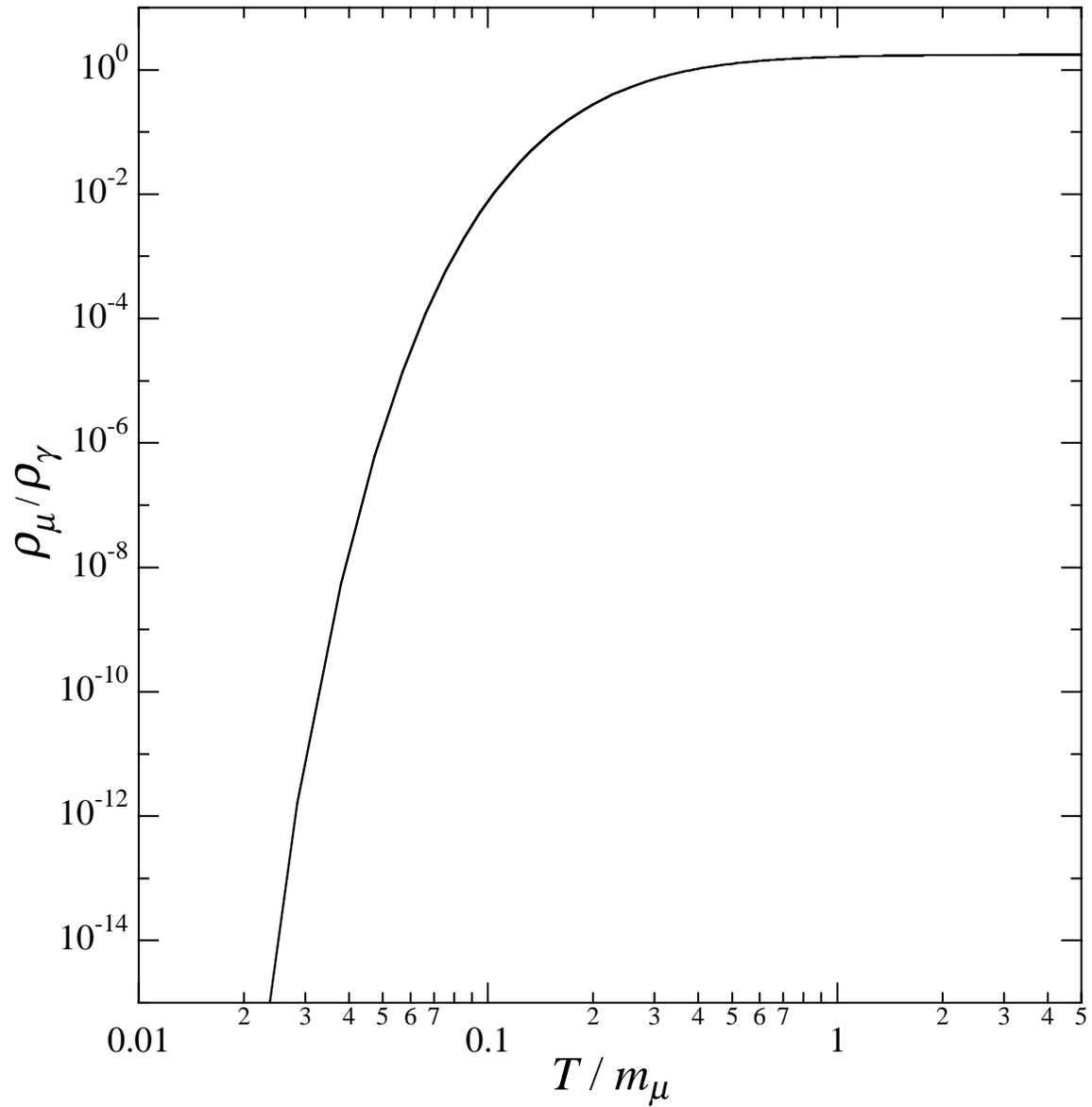}
\caption{Ratio of the energy density of muons (or any massive particle $\mu$)
to photons as a function of the ratio of temperature to rest mass.
This shows that even at a temperature of only $\sim 20\%$ of the
rest mass, a significant fraction ($\sim 10\%$) of the energy density 
still resides in particle/antiparticle pairs.}  
\protect \label{fig:2}
\end{figure}

One does not need to increase the  decoupling temperature by much before
photon heating by annihilations becomes important for determining 
the relic neutrino temperature.  For illustration,
Figure \ref{fig:2} shows the ratio of muon to photon energy
densities as a function of temperature in units of the muon rest
mass $m_\mu = 105 $ MeV.  A similar curve could be drawn for
any massive species.
One can see that even at a temperature
of only 20\% of the muon rest mass,  muons still contribute
about 10\% of the mass energy density, and hence, can affect
the ratio of the  photon to neutrino 
temperatures as these remaining muons annihilate.  
 Combining Figures \ref{fig:1} and \ref{fig:2}, one can see
 that even for a degeneracy
parameter of $\xi_\nu \sim 6$, the decoupling temperature
is at 20\% of $m_\mu $.  
For the case of highly degenerate neutrinos ($\xinum 
%
\ga 10.8$
and  $\xinut 
%
\ga 9.8$), $T_D(\xi_{\nu_i})$ can exceed the muon
rest energy  and even the QCD phase transition temperature. 
 
If the neutrinos  decouple early, they are
not heated as the number of particle degrees of freedom change.
Hence, the ratio of the neutrino to
photon temperatures, $T_\nu/T_\gamma$, is reduced.
The computation of the ratio of the final present neutrino
temperature to the photon temperature is straightforward.
Basically, since the universe is a closed system,
the relativistic entropy is conserved  within
a comoving volume. 
That is;
\begin{equation}
R^3 s = {~\rm~Constant}~~,
\end{equation}
where the entropy density $s$ is defined,
\begin{equation} 
s \equiv \sum {(\rho_i + p_i - \mu_i n_i) \over T} 
= {2 \pi^2 \over 45} g_{eff} T^3 ~~,
\label{sdef}
\end{equation}
and the sum is over all species present.  Since the mass-energy
is dominated by relativistic particles $s$ can be written in terms
of the effective number of particle degrees of freedom,
\begin{equation}
g_{eff} = \sum_{i=bosons} g_i \biggl({T_i \over T}\biggr)^3
 + {7\over 8} \sum_{i=fermions} g_i \biggl({T_i \over T}\biggr)^3~~.
\label{geff}
\end{equation}
As each species annihilates,
$s R^3$ remains constant.  Therefore, 
 the temperature of the remaining species
increases by a factor of 
$(g_{eff}^{before}/g_{eff}^{after})^{1/3}$. 
This accounts for the
usual heating of photons relative to neutrinos
due to $e^\pm$ pair annihilations by a factor of $(11/4)^{1/3}$.
Note, that in the computation of $g_{eff}$ from equations (\ref{sdef})
and (\ref{geff}) it is important to evaluate the energy densities
continuously (cf. Figure \ref{fig:2}) and not simply  assume
abrupt annihilation as the temperature approaches
the rest energy of each particle as was done in Kang \& Steigman (1992).

\begin{figure}
\plotone{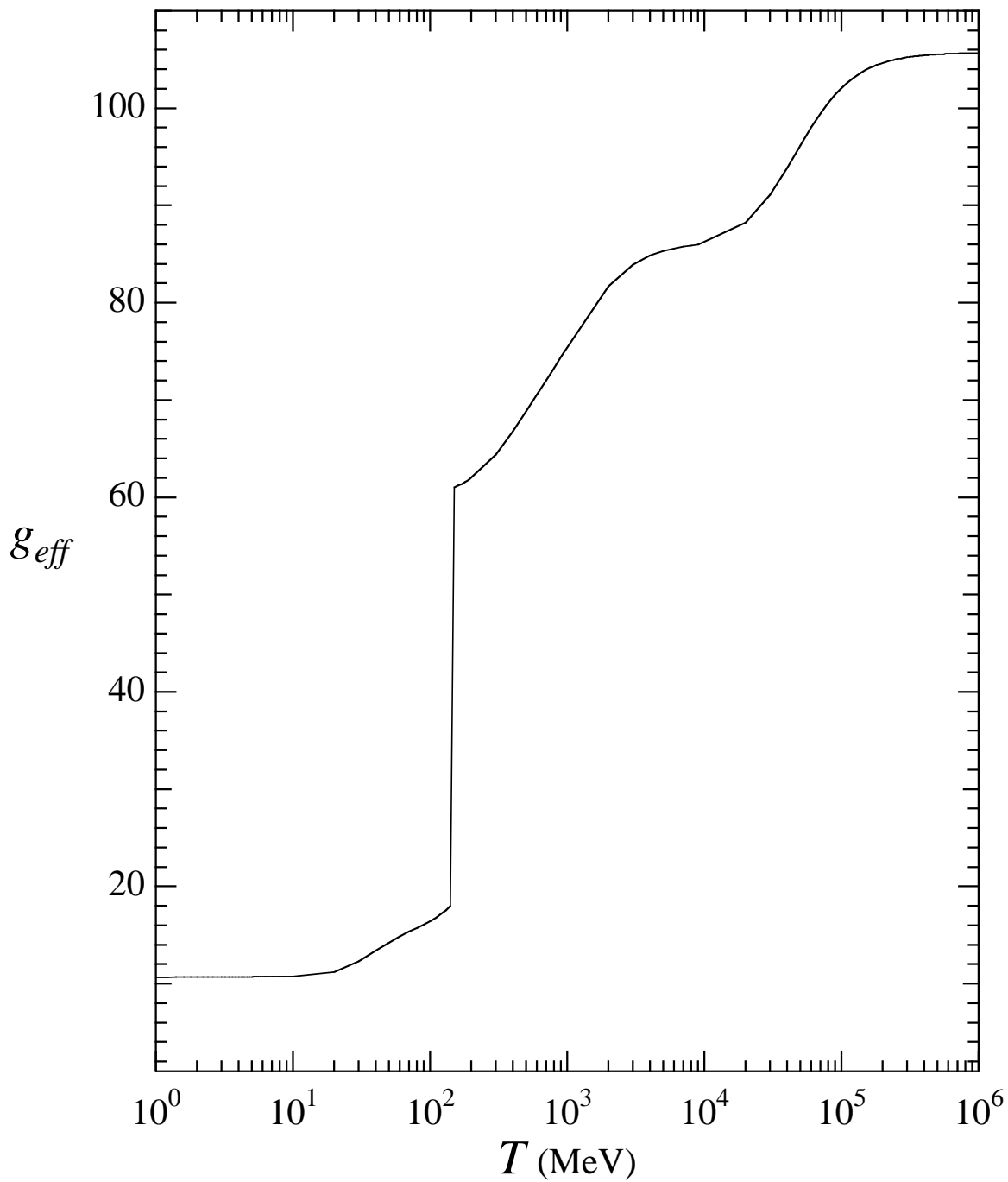}
\caption{An illustration of the effective degrees of freedom $g_{eff}$ as a function of
temperature for a standard big bang with 3 nondegenerate neutrinos.
For this illustration we have assumed that the neutrinos do not decouple
until $T \approx$ MeV.   The discontinuity at $T = 150$ MeV is due to the
QCD phase transition.}
\protect\label{fig:3}
\end{figure}

Figure \ref{fig:3} illustrates  $g_{eff}$ as a function of temperature
from 1 MeV to 1 TeV. In each calculation $g_{eff}$ depends upon the
relic neutrino temperature and therefore when the neutrinos decoupled.
For purposes of illustration only, we drawn this figure for
three neutrino types which do not decouple until $\approx$ MeV.
To construct this figure we have included:
mesons ($\pi$, $\rho$, $\phi$, $\omega$, $\eta, \eta'$);
leptons ($e$, $\mu$, $\tau$, $\nu_e$, $\nu_\mu$, $\nu_\tau$);
a QCD phase transition at 150 MeV;
light quarks and gluons ($u, d, gluon$);
s-quarks (with $m_s = 120$ MeV);
c-quarks (with $m_c = 1200$ MeV);
b-quarks (with $m_b =$  4250 MeV);
W, Z-bosons (with $m_Z$ = 80 GeV);
and t-quarks (with $m_t =$ 173 GeV).

One can see from this figure that if neutrinos decouple 
before the QCD phase transition, there is substantial 
 heating of photons due to the large change in degrees of freedom
at this epoch.
%
%

\begin{figure}
\plotone{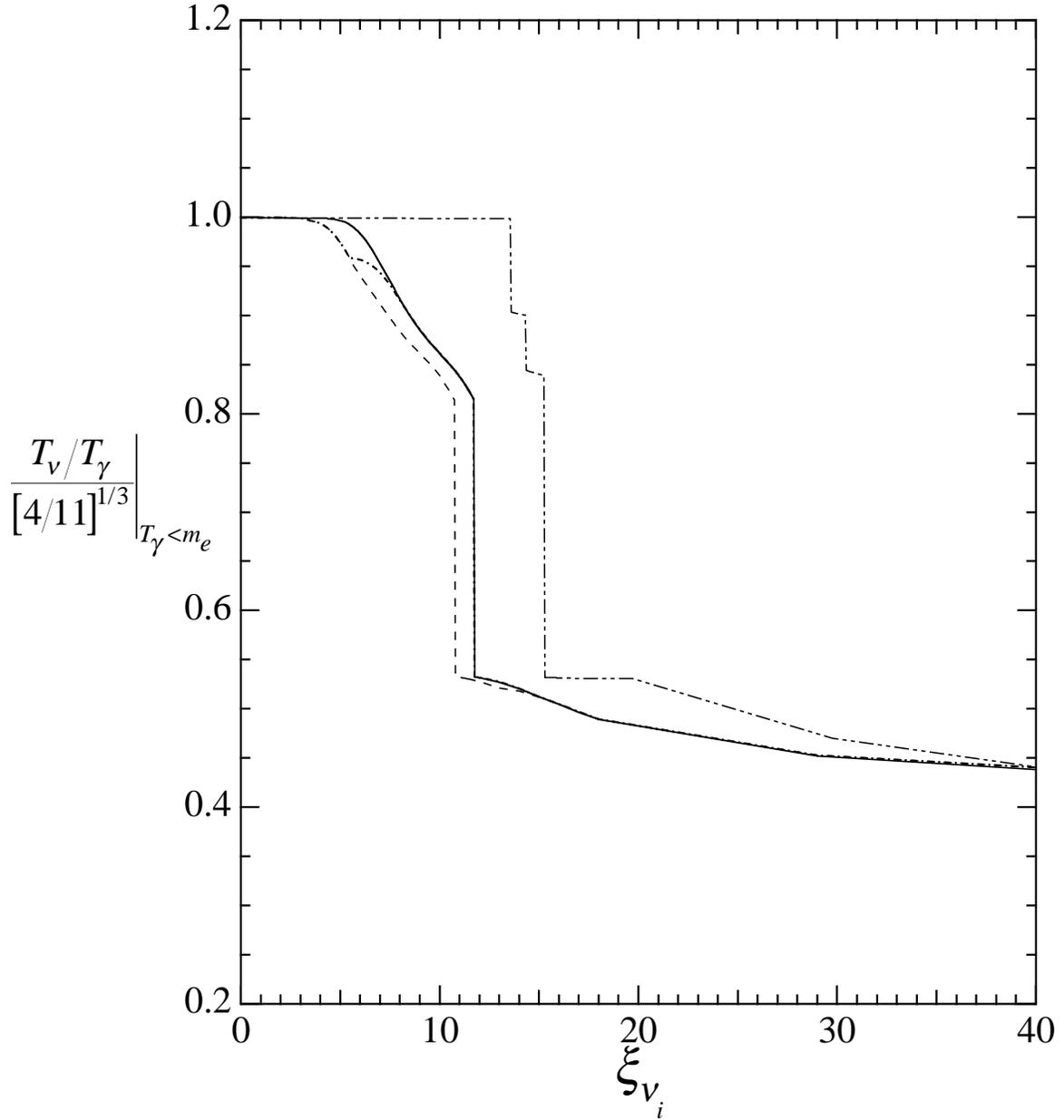}
\caption{Ratio of relic neutrino temperature to photon temperature
as a function of degeneracy parameter for each neutrino species for 
$\nu_e$ (solid line), $\nu_\mu$ (dot-dashed line), $\nu_\tau$ (dashed
line), compared with the previous 
estimate (two-dot dashed line) of Kang \& Steigman (1992)}
\protect\label{fig:4}
\end{figure}

Figure \ref{fig:4} shows the final ratio of neutrino temperature today
to that of the standard non-degenerate big bang for three neutrino flavors.
The biggest drop in temperature for all three neutrino flavors
occurs for $\xi_\nu \approx 10$.  This corresponds to a decoupling
temperature above the cosmic QCD phase transition.  The low temperature is
the result of the decrease in particle 
degrees of freedom during this phase transition.
This discontinuity will have important consequences in the
subsequent discussions.

For comparison, the same computation from Figure 1 of 
Kang \& Steigman (1992) is shown.
One can see that there is a substantial difference between the two results.
These differences  stem from two effects.  On the one hand the decoupling
temperature is higher in the present work for a given $\xi_{\nu_i}$ because
of the various corrections as explained above.
 This is the
reason that the discontinuity in relic neutrino temperature occurs 
at $\xinumt \approx 11$ instead of 15 as in Kang \& Steigman (1992).
The second difference is that here we have calculated the proper 
thermodynamic energy density
for each fermion and boson species continuously (cf. Eq. 1), whereas in
Kang \& Steigman (1992) an approximation was made that each particle species
instantaneously annihilates as the temperature drops below its rest energy.
This is the reason for the step function appearance of the curve
in Kang \& Steigman (1992).

Some discussion of book keeping here is in order as it explains why
we observe the jump in relic temperature when neutrinos decouple
above the QCD transition.  
Let us consider a neutrino which decouples just
above the QCD transition. Then the relevant  fermion degrees
of freedom consist mainly of 3 $q-\bar q$ colors (treating the $s$-quark
as relativistic)  $\times$ 3 flavors contributing
$g_{f} = 31.5$, plus  $e^+-e^-$ and  $\mu^+-\mu^-$ plus
2 remaining nondecoupled flavors of $\nu-\bar \nu$ lepton pairs.
These  contributing another $g_{f} = 10.5$.
For the bosons there are 8 gluons and 1 photon contributing a
total $g_{b} = 18$.  While the species are in equilibrium the total
$g_{eff} = 60$.  If  no more species decouple before the electron pair
annihilation epoch, the 
fermion degrees of freedom  before electron-positron annihilation
(at $T \approx 1 $ MeV) include only 2 flavors of $\nu-\bar \nu$
plus  $e^+-e^-$ so that $g_{f} = 7$,
while for bosons 
only photons remain  so that $g_{b} = 2$ and $g_{eff} = 9$.
The relic neutrino temperature is therefore reduced  by 
an additional factor due to these differences
in particle degrees of freedom $(T_\nu/T_\gamma)/[4/11]^{1/3} = (9/60)^{1/3}
\approx 0.53$ (cf.~Figure \ref{fig:4}).

For all three neutrino flavors the temperature begins to
decrease relative to the standard value for a degeneracy parameter
as small as $\xi \sim 5$.  This is because some relic $\mu -  \bar \mu$
pairs are still present even at temperatures well below the muon rest energy 
(cf. Figure \ref{fig:2}).  The first  neutrino species
to be affected as the degeneracy increases are the $\nu_\mu$ and
$\nu_\tau$. They decouple at a higher temperature than $\nu_e$
even in the standard nondegenerate big bang because the electrons
continue to experience charged-current interactions to lower temperature.  
The muon neutrinos exhibit a slightly different
behavior than $\nu_\tau$ for degeneracy parameters $\xinum > 5$ because the
$\mu -  \bar \mu$ density
is large enough at the decoupling temperature for charged-current
interactions to be significant.  This maintains equilibrium
for $\nu_\mu$  to somewhat
lower temperatures even for degenerate neutrinos.  This causes the
$\nu_\mu$  decoupling temperature to be lower (cf.~Figure \ref{fig:1})
and the relic temperature  to be slightly higher than the 
$\nu_\tau$ temperature for degeneracy parameters between
5 and 9.

\section{PRIMORDIAL NUCLEOSYNTHESIS}
\subsection{Current Status}
Although the homogeneous BBN model has provided strong support for the
standard hot big-bang cosmology,
possible conflicts have emerged between the predictions of
BBN abundances as the astronomical data
have become more precise in recent years.  One difficulty has
been imposed by recent detections of a low deuterium abundance
(\cite{burles98a}; \cite{burles98b}, see also \cite{levshakov99}) 
in Lyman-$\alpha$ absorption systems along the line of sight to high
red-shift quasars.  The low $\dovh$ favors a high baryon content universe
and a high primordial $\he4$ abundance, $Y_p \ga 0.244$.  This
is inconsistent with at least some of the reported 
constraints from measurements of a low
primordial abundance of $\he4$, $Y_p \approx 0.235 \pm 0.003$,
in low-metallicity extragalactic H II regions
(\cite{olivestei95}; \cite{steigman96};
\cite{hata96}; \cite{olive97}; \cite{kajino98a}; \cite{piembert00}).
This situation is exacerbated by recent detailed analyses 
(\cite{esposito99}; Lopez \& Turner 1999) of the 
theoretical uncertainties in the
weak interactions affecting the neutron to proton ratio at the onset of
primordial nucleosynthesis.
These results require a positive net correction to
 the theoretically determined $\he4$ mass fraction $Y_p$ 
of $+0.004$ to  $+0.005$ or $\sim$2\%. 
We also note that the low deuterium abundance is 
marginally inconsistent with the
 $\li7$ abundance inferred by measurements
of lithium in Population II  halo stars (\cite{ryan99}; \cite{kajino00}).
Significant depletion of lithium from these stars, or  a lower reaction
rate for primordial lithium production may be required.

Another potential difficulty has been imposed by recent 
X-ray observations of
rich clusters (\cite{white93}; \cite{white95}; \cite{david95};
\cite{bahcall95}).  The implied baryonic contribution to the closure 
density is $0.08 \le \omegab h_{50}^{3/2}/ \Omega_M \le 0.22$
(\cite{tytler00}),  where $\Omega_M$ is
the total matter (dark plus visible) contribution, and $h_{50}$ is the 
Hubble constant $H_0$ in units of 50 km~s$^{-1}$ Mpc$^{-1}$.
Consistency with the limits ($0.03 \le \omegab\h502 \le 0.06$) 
from homogeneous BBN (\cite{walker91}; \cite{smith93};
\cite{copi95}; \cite{ScMa95}; \cite{osw99}) then requires that
$0.14 \le \Omega_M h_{50}^{1/2} \le 0.75$.  Hence,  
matter dominated cosmological models  (for example with $H_0 = 75$
km~s$^{-1}$ Mpc$^{-1}$  and $\Omega_M \ge 0.61$) can be in conflict with BBN.

Another possible conflict has emerged in recent measurements of
the power spectrum of the cosmic microwave background (CMB)
temperature fluctuations.  A suppression of the second acoustic peak in the power
spectrum has been recently reported  in the
balloon based CMB sky maps from the BOOMERANG (\cite{boomerang})
and MAXIMA-1 (\cite{hanany}) collaborations.
Indeed, the derivation of cosmological parameters
from these new data sets (\cite{lange}; \cite{balbi})  indicate
a preference for a large baryon density.  For example,
the  optimum cosmological models consistent with the BOOMERANG
data  (\cite{lange}) indicate  $\Omega_b h^2 = 0.032 \pm 0.004$ $(1 \sigma)$
 for fits in which there is no prior restriction on the range of $\Omega_b h^2$.
Similarly, the likelihood analysis based upon the MAXIMA-1 data
(\cite{balbi}) indicates $\Omega_b h^2 = 0.025 \pm 0.005$.  These
results are to
be compared with the best current $(1 \sigma)$ limit from
standard primordial nucleosynthesis without neutrino degeneracy,
$\Omega_b h^2 = 0.019 \pm 0.0024$ (\cite{osw99};\cite{tytler00}) which
come from the deuterium abundance in Lyman-$\alpha$ clouds observed
along the line of sight to background quasars.
The  CMB data indicate a baryon content which is at least 1-2
$\sigma$ above the optimum  BBN value. Indeed,
$\Omega_b h^2 \ge 0.20$ would require a
primordial helium abundance of $Y_p \ge 0.25$ which is  beyond even
the most generous adopted limits from observation (\cite{osw99}).
Thus, it seems that that the CMB data hint at (though not necessarily  prove)
the need for a modification of BBN which allows higher values
of $\Omega_b h^2$ while still satisfying the constraints from
light element abundances.  Such conditions are easily satisfied
by neutrino-degenerate models.

\subsection{Neutrino-Degenerate BBN}
Most previous works have only considered the effects of 
neutrino degeneracy on  the light elements $\he4$, D, and $\li7$.
Recently, the predicted abundance of other elements such as $\li6$, $\be9$,
and $\b11$ in a neutrino-degenerate universe were  also studied
(\cite{kim95}; \cite{kim98}). Here we investigate the effects
of lepton asymmetry on the predicted abundances of heavier elements
($12 \le A \le 18$) as well as these light elements.  

Non-zero lepton
numbers primarily affect nucleosynthesis in two
ways (\cite{wagoner67}; \cite{terasawa85}; \cite{terasawa88};
\cite{scherrer83}; \cite{yahil76};
\cite{beaudet76}; \cite{beaudet77}; \cite{david80};
\cite{olive91}; \cite{kang92}; \cite{starkman92}; \cite{kajino98a};
\cite{kim95}; \cite{kim98}). First, degeneracy in any neutrino species
leads to an increased universal expansion rate independently of
the sign of~$\xinue$ (cf.~Eqs.~\ref{dense} and \ref{friedmann}).
As a result, the weak
interactions that maintain neutrons and protons in
statistical equilibrium decouple earlier.
This effect alone would lead
to an enhanced neutron-to-proton ratio at the onset of the
nucleosynthesis epoch
and increased $\he4$ production.

Secondly, a non-zero electron neutrino
degeneracy can directly affect the equilibrium n/p ratio at weak-reaction
 freeze out. The equilibrium n/p ratio is related to the electron neutrino
degeneracy by $\rm{n/p} = exp\{-(\Delta \it{M/T}_{n \leftrightarrow p}) -
\xinue\}$,
where $\Delta M$ is the neutron-proton mass difference and
$T_{n \leftrightarrow p}$ is the
freeze-out temperature for the weak reactions
converting protons to neutrons and vice versa.
This effect leads to either increased or decreased $\he4$ production,
depending upon the sign of $L_e$ or $\xinue$.

There is also a third effect which we emphasize 
in this paper. As discussed in the previous section,
$T_\nu/T_\gamma$ can be reduced if the neutrinos decouple
at an earlier epoch.  This lower temperature
reduces the energy density of the highly degenerate
neutrinos during the BBN era, and hence, slows down the expansion of the
universe. This leads to decreased $\he4$ production.
We show in the next
section that the allowed values for $\xinue, \xinum, \xinut$ and $\omegab$
which satisfy
the light-element abundance constraints  are significantly changed for large
degeneracy ($\xinum,~\xinut \ga 10$) compared to the results of 
previous studies.

\subsection{Summary of Light-Element Constraints}

The primordial light element abundances
deduced from observations have been reviewed in a number
of recent papers (\cite{osw99}:
\cite{nolett00}; \cite{steigman00}; \cite{tytler00}).
There are several outstanding uncertainties.  For primordial helium
there is an uncertainty due to the fact that deduced abundances tend
to fall in one of  two possible values, one high and the other low.
Hence, for \he4 we adopt a wide range:
	 $$ 0.226 \leq \yp \leq 0.247.  $$
For deuterium there is a similar possibility for either a
high or low value.  Here, however, we adopt the generally accepted
low values of Tytler et al.~(2000),
	$$ 2.9 \times 10^{-5} \leq \dovh \leq 4.0 \times 10^{-5}.  $$
For primordial lithium there is some uncertainty from the possibility
that old halo stars may have gradually depleted their primordial
lithium.  Because of this possibility we adopt the
somewhat conservative constraint:
	$$ 1.26 \times 10^{-10} \leq \li7/{\rm H} \leq
	 3.5 \times 10^{-10}$$

\subsection{Nucleosynthesis Model}

As we shall see, shifts in the relic neutrino
temperature can dramatically affect the
abundance yields  (\cite{kajino98b}).
We now explore the parameter space of neutrino
degeneracy and baryonic content to reinvestigate the
range of models compatible with the constraints from light
element abundances.  

  For the present work we have applied a standard big bang code
with all reactions updated up to A=18.  
[However, in the present discussion only reactions involving
nuclei up to A=15 are significant.]  
In this way possible effects of lepton asymmetry on 
heavier element abundances could be analyzed along with the
light elements.  
In this context a recent compilation of the nuclear reaction 
rates relevant to the production of
$\b11$ (\cite{orito98}) was useful because several important LiBeB(a,x)
and (n,$\gamma$) reaction rates in the 
literature sometimes differ from one another
by 2-3 orders of magnitude.  The calculated abundances of 
heavier elements based upon these rates can
also differ from one another by 1-2 orders of
magnitude. We carry out BBN calculations which include all of
the recent compilations of reaction rates relevant to the production of
isotopes  including those that are heavier than $\li7$ up to $^{18}$O 
(i.e.~\cite{orito98}; \cite{mohr99}; \cite{angulo99};
\cite{herndl99}; \cite{wagemans99}; 
and any other previous estimates are considered).

\subsection{Neutrino Degeneracy Parameters}
We have explored a broad range 
of the parameter space of neutrino-degenerate models.
The main effects of the inclusion of either  \num~ or \nut~
degeneracy on BBN is an enhancement of energy density of the universe.  
The values for $\xinum $ and $\xinut$ primarily affect the expansion rate.
This means that $\xinum$
and $\xinut$ are roughly interchangeable as far as their effects on 
nucleosynthesis are concerned.  Furthermore, we expect that the
net total lepton number is small though the lepton number for individual
species could be large.  Hence, it is perhaps most plausible 
to assume that
the absolute values of $\vert \xinum\vert$ and $\vert \xinut\vert $
are nearly equal.  Therefore, in what follows, we describe results
for $\vert \xinum\vert =\vert \xinut\vert \equiv \xinumt$. 
This reduces the parameter space to
three quantities: $\omegab$, $\xinue$, and 
$\vert \xinum\vert =\vert \xinut\vert$.
The calculations were conducted by first choosing a value
for $\Omega_b \h502$. It was then necessary
to find a value of $\xi_{\nu_e}$ for which the light element constraints 
could be satisfied for some value of $\xinumt$.

\subsection{Results}
\begin{figure}
\plotone{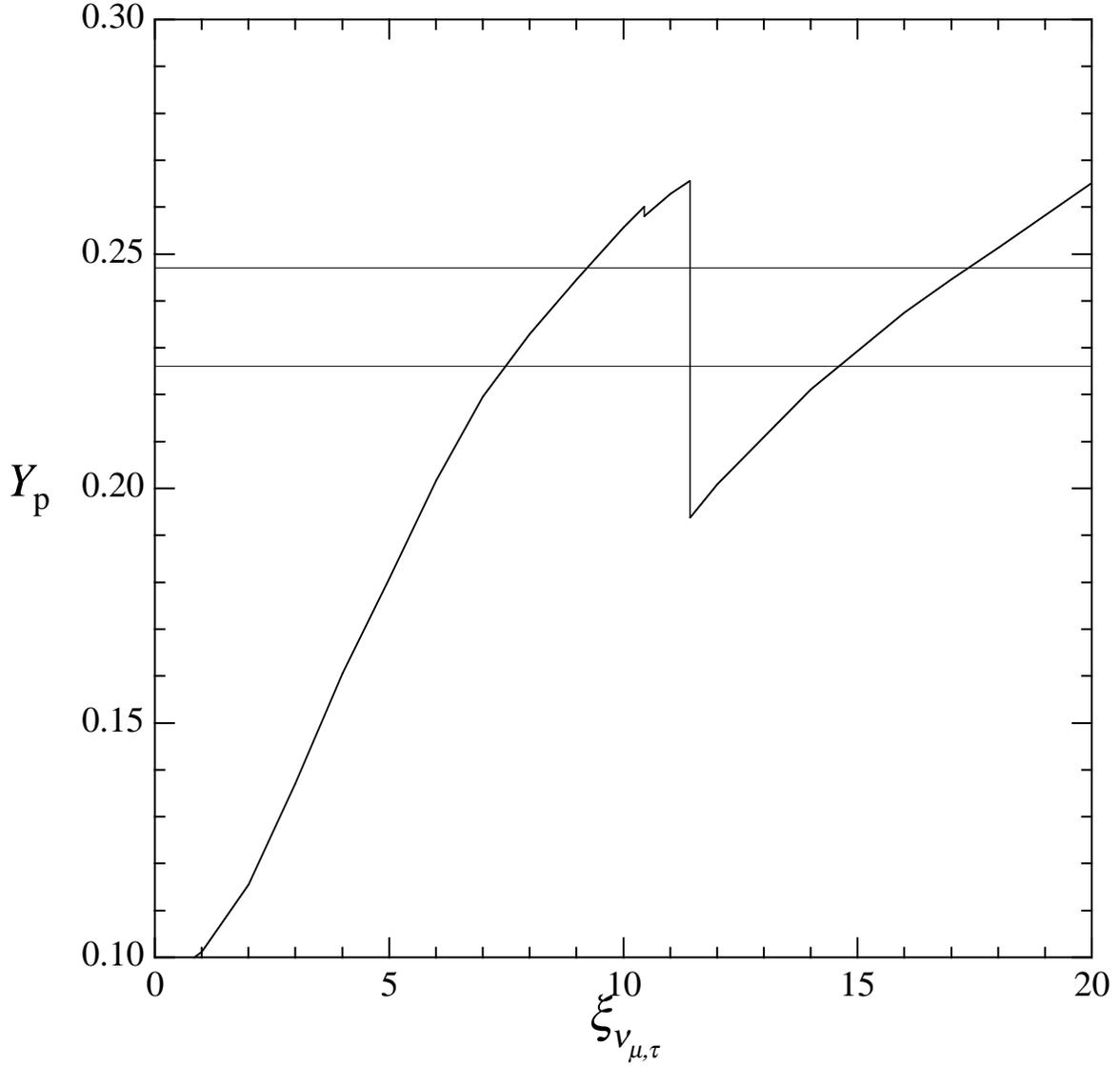}
\caption{
Helium mass fraction $Y_p$ as a function of degeneracy
parameter $\xinumt$ for
$\omegab \h502 = 0.3$ and $\xinue = 0.9$.
Horizontal lines show the adopted observational constraints.}
\protect\label{fig:5}
\end{figure}

As an illustration, Figure~\ref{fig:5} shows a calculations of the
helium abundance as a function
of $\xinumt$ in a model with $\Omega_b \h502 =
0.3$.  It was found in this model that the helium constraint could be satisfied
 for $\xi_{\nu_e}$ in the range of 0.79 to 0.94.  This figure
is for $\xinue = 0.9$.  Our results for
$\xinumt  < 10$ are consistent with those of
Kang \& Steigman (1992) if we run for a similar baryon-to-photon
ratio $\eta_{10} = 5$ or $100$ and utilize the 
neutron half life adopted in that paper.  

\begin{figure}
\plotone{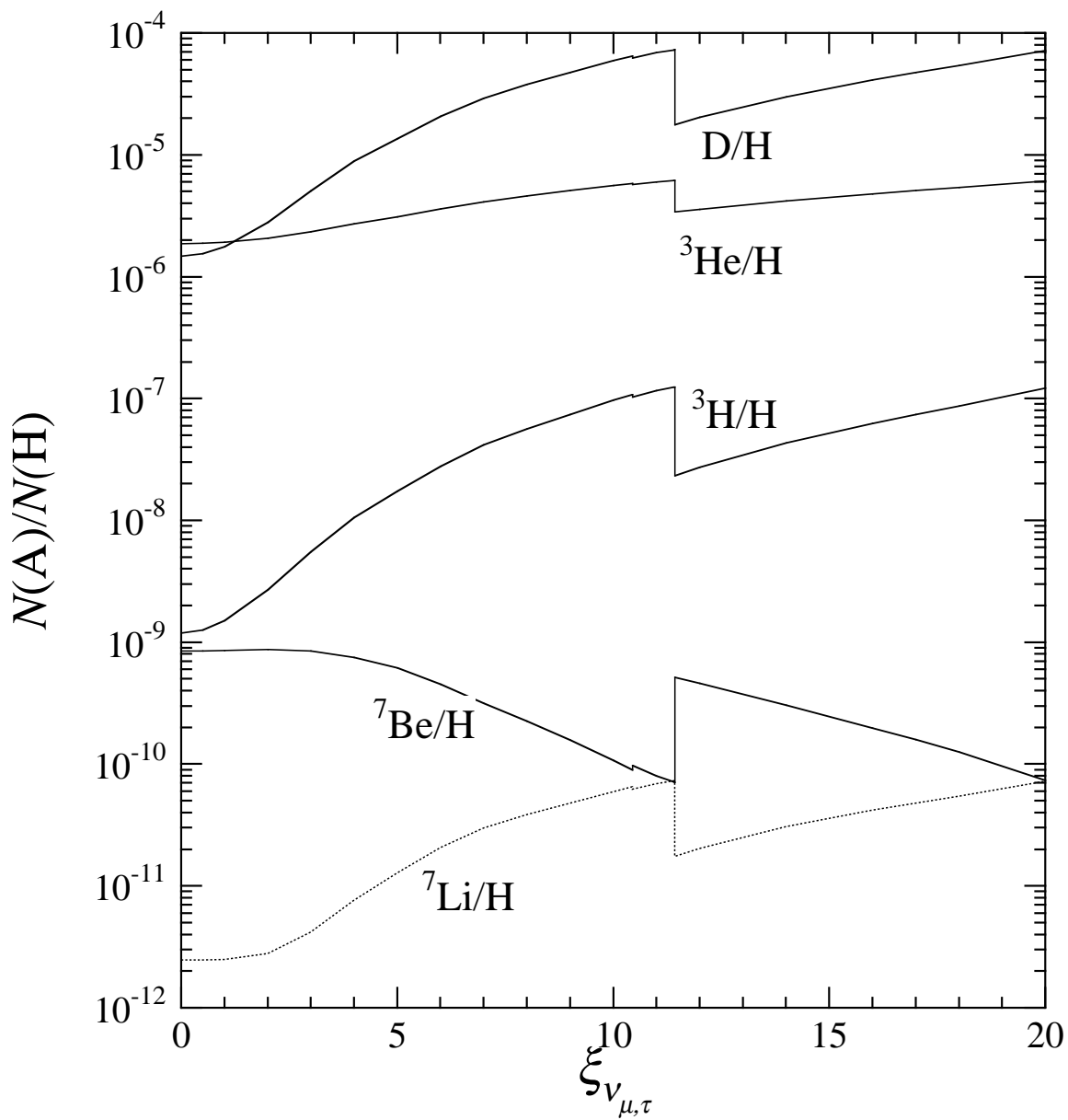}
\caption{The predicted abundances of the light elements as a function of the
neutrino degeneracy $\xinumt$ for $\omegab \h502$ = 0.3 and $\xinue = 0.9$.}
\protect\label{fig:6}
\end{figure}

In our calculations  for helium (Figure \ref{fig:5}) and other light nuclides
(Figure \ref{fig:6}) we see two different jumps in the
nucleosynthesis yields.  There is  a small one for $\xinumt \approx 10$
and a large one for $\xinumt \approx 11.4$.   These two jumps
are due to the fact
that we consider two degenerate neutrino species which
decouple at slightly different temperatures (due to the presence
of muons to interact with $\nu_\mu$).  Therefore,
one species ($\nu_\tau$) decouples above the QCD transition
for a smaller value for  $\xinumt$. However, this jump has a smaller
effect on nucleosynthesis because so many degrees of freedom are still carried
by the other degenerate neutrino species ($\nu_\mu$) which decouples later.

The main difference between our results and
Kang \& Steigman (1992), however, is that the helium
abundance  is significantly changed at high values of $\xinumt$.
This results from our higher decoupling temperatures
and therefore the lower relic neutrino temperatures 
during primordial nucleosynthesis  (See Figure \ref{fig:4}).
This effect of varying $T_\nu/T_\gamma$ for large $\xi_\nu$ on BBN has 
not been adequately explored in the previous studies.

Figure~\ref{fig:6} illustrates the effects on the other
light-element abundances for this particular parameter set.  
This figure shows that  for moderate values of $\xinumt$
the main effect is that weak reaction
freeze-out occurs at a higher temperature.  The resultant
enhanced n/p ratio increases the
abundances of the neutron-rich  light
elements, D, $^{3}\rm{H}$ and $\li7$, while the $\be7$ abundance 
decreases.  

 Regarding $\li7$ and $\be7$, the enhanced 
expansion rate from neutrino degeneracy affects the
yield of $A = 7$ elements in two  different ways. These elements 
are produced mainly by
the nuclear electromagnetic capture reactions:  $t(\alpha,
\gamma)\li7$ and $\he3(\alpha, \gamma)\be7$. 
Hence, the production of these elements
begins at a later time and lower temperature
than the other light elements because they require 
time for a significant
build up of the reacting A = 3 and 4 elements. In
a neutrino-degenerate universe, however,
the increased expansion rate, hastens the freeze-out of these
reactions from nuclear statistical equilibrium (NSE) 
resulting in reduced $A=7$ yields relative to the
nondegenerate case. However, the enhanced n/p ratio also 
increases the
tritium abundance in NSE.  This effect tends to
offset the effect  of rapid 
expansion on the production of $\li7$.  
The net result is more $\li7$ production.

As in the  case of primordial helium,  there is a rapid
change in the nucleosynthesis yields on Figure \ref{fig:6}
 once the $\nu_\mu$ and $\nu_\tau$
 decoupling temperatures separately 
exceed the epoch of the QCD phase transition.  The ensuing
lower neutrino temperature during primordial nucleosynthesis
then resets the abundances to those of a lower effective degeneracy.
The two discontinuities in Figure~\ref{fig:6} 
at $\xinumt = 10$ and 11.4 correspond to
the points at which respectively 
the tau  or muon neutrino decoupling temperatures
separately exceed the QCD phase transition temperature
(cf.~Figure \ref{fig:4}).

Figure~\ref{fig:7a} summarizes the allowed regions of the
$\xinue$ vs.~$ \xinumt$ plane based upon the various indicated
light element constraints in a universe with $\omegab \h502 = 0.1$.  
The usually identified allowed region (cf.~\cite{kohri97}) for 
small $\xinue \sim 0.2$ and
$\xinumt \sim 2$ is apparent. Indeed, it has been
argued (cf. \cite{kohri97}) that such degeneracy
may be essential to explain the differences in the constraints
from primordial helium and deuterium.

Figures \ref{fig:7b} and 
\ref{fig:7c} show the same plots, but for 
$\omegab \h502 = 0.2$ and $\omegab \h502 = 0.3$, respectively.
The decline in the primordial deuterium abundance
for models in which $\xinumt > 10$ allows for new regions 
of the parameter space in which the light element constraints 
can be accommodated.  This new allowed region for
large degeneracy persists as the baryon density is increased.

Figure \ref{fig:8} highlights the basic result of this study.
It shows that there exists a single parameter
(either $\xinumt$ or $\omegab \h502$) family of neutrino degenerate
models allowed by BBN.
For low $\omegab \h502 $ models, only the usual low values
for $\xinue $ and $\xinumt$ are allowed.  Between
$\omegab \h502 \approx$ 0.187 and 0.3, however, more
than one allowed region emerges.
For $\omegab \h502 \ga 0.4$ only the large degeneracy solution
is allowed.  Neutrino degeneracy can even allow baryonic densities
up to $\omegab \h502 = 1$.  This result
has been noted previously (cf. \cite{kang92}: \cite{starkman92}).  
What is different here is that
the high $\omegab \h502$ models are made possible for smaller
values of $\xinue $ due to the higher neutrino decoupling
temperatures deduced in the present work.
As we shall see, this allows the circumvention of other 
cosmological constraints as well suggesting
that baryons and degenerate neutrinos
might provide a larger contribution to the
universal closure density than has previously been derived.

Figures \ref{fig:9}-\ref{fig:11} illustrate the elemental
abundances obtained in  the family of
allowed models as a function of $\omegab \h502$.
Figure \ref{fig:11} in particular allows  us to 
consider whether there exists an abundance signature 
in other elements which might
distinguish this new degenerate neutrino solution
from standard BBN.
For the most part the yields of the light and heavy 
species are similar to
those of the standard non-degenerate big bang.  However, the
boron abundance exhibits an increase with baryon density due
to alpha captures on $^7$Li. 
Thus, in principle, 
an anomalously high boron abundance together with 
beryllium and $^6$Li
similar to that expected from
standard BBN might be a signature
of neutrino-degeneracy.

\section{OTHER COSMOLOGICAL CONSTRAINTS}
  We have seen that a new parameter space in the constraints from
light elements on BBN emerges in neutrino-degenerate
models just from the fact that the relic neutrino temperature
is substantially diminished when degeneracy pushes neutrino decoupling to
an earlier epoch.  The viability of this solution requires
large neutrino degeneracy. Hence, it becomes necessary to reexamine constraints
posed from other cosmological considerations.

\subsection{Structure Formation}
It has been argued (\cite{kang92})
 that large neutrino degeneracy
is ruled out from the implied delay in galaxy formation in
such a hot dark matter universe.  This argument is summarized
(\cite{kang92}) as follows:
  
Neutrino degeneracy speeds up the expansion rate by a factor 
\begin{equation}
S_0^2(\xi_\nu)  = 1 +  0.135 (F(\xi_\nu) - 3) \approx 0.135 
F(\xi_\nu)~,
\end{equation}
 where $F(\xi_{\nu})$ is an effective energy density factor (\cite{kang92})
for neutrinos.  For massless neutrinos we have from Equations \ref{dense}
and \ref{friedmann},
\begin{equation}
F(\xi_{\nu}) =  \sum_i F(\xi_{\nu_i}) = \sum_i \biggl({T_{\nu_{i}} \over T_\gamma} \biggr)_{Nuc}^4
 \biggl[1 + {15 \over 7} \biggl(
{\xi_{\nu_i} \over \pi}\biggr)^4 + {30 \over 7} \biggl(
{\xi_{\nu_i} \over \pi}\biggr)^2\biggr]~,
\end{equation}
where $(T_{\nu_{i}} /T_\gamma )_{Nuc}$ is the neutrino-to-photon
temperature ratio before $e^\pm$ pair annihilation. 
This ratio is unity in the standard model with  little or no degeneracy. 

With the increased mass-energy in degenerate neutrinos, 
the time of matter-radiation equality
occurs at smaller redshift and there is less time for the
subsequent growth of fluctuations. The redshift $z_{eq}(\xi_{\nu})$
for matter-radiation equality for a neutrino-degenerate 
universe can be written in terms of the redshift for
a nondegenerate universe and the speed-up factor,
\begin{equation}
1 + z_{eq}(\xi_{\nu}) = S_0^{-2}(1 + z_{eq}(\xi_{\nu}=0))~,
\end{equation}
 For the present matter-dominated universe without neutrino degeneracy
and $\Omega_{M} \le 1$  we have
$1 + z_{eq}(\xi_{\nu}=0) \la 1.06 \times 10^4$ (\cite{kang92}).
Furthermore, one demands that
the fluctuation amplitude  $A(z)$
grows at least linearly  with redshift,
i.e.~$A(z) > (1 + z_{eq}) A(z_{eq})$.  One also
requires that the amplitude at least reaches 
unity by the present time,
\begin{equation}
{ 1 \over  A(z_{eq})} \la (1 + z_{eq}) \la {A(z = 0) \over A(z_{eq})} ~.
\end{equation}
  Thus, the requirement
of sufficient growth in initial perturbations places a bound on
the speed-up.  Namely,
\begin{equation}
S_0^2 \la {1.06 \times 10^4 \over 1 + z_{eq}(\xi_{\nu})}
 \la 1.06 \times 10^4 A(z_{eq})~.
\end{equation}
Then requiring that
$A(z_{eq}) \la 10^{-3}$  (\cite{steigman85}) 
leads to the constraint that $F(\xi_{\nu}) \la 10.6/.135 \approx 79$.  

Figure \ref{fig:13} shows the $F(\xi_{\nu})$ and
$F(\xi_{\nu_{i}})$ calculated as a function of $\xinumt$
for the allowed models of Figure \ref{fig:8}.
Since our interesting parameter regions in Figure \ref{fig:8}
satisfy $\xinue \ll \xinumt$, only $\nu_{\mu,\tau}$ contribute 
significantly to the total $F(\xi_{\nu})$.  
Also shown are values
of $F(\xi_{\nu_{i}})$ if only one neutrino species was degenerate.
The dashed line gives the constraint $F(\xi_{\nu}) \la 79$ 
from Kang \& Steigman (1992).
For comparison, the two-dot dashed line also shows the 
$F(\xi_{\nu})$ for a single species from Kang \& Steigman (1992).
For low values of $\xinumt \la 4$, our $F(\xi_{\nu_i})$ values
for a single degenerate species are  nearly identical 
to those of Kang \& Steigman (1992).
However, as $\xi_{\nu_{i}}$ increases, 
our curves are lower due to the fact that we treat
the annihilation epochs continuously (cf.~Figure \ref{fig:4})
rather than discretely, except for the QCD transition.  
By chance, our limit in the low
degeneracy range for two degenerate neutrinos,
($\xinumt \la 6.2$) is close to the single species
limit ($\xi_i^{KS} \la 6.9$) of Kang \& Steigman (1992). 
In the present work, however, the single species limit
for moderate degeneracy increases to $\xi_i \la 8.2$.
Moreover, in the present work
we now find that there also exists 
a new allowed region  with $11.4 \la \xinumt \la 13$ for which
this growth constraint is satisfied even for two degenerate
species.  For only one degenerate species the  new allowed range
expands to $11.4 \la \xinum \la 16$.  An upper limit
of $\xinumt \la 13$ for two degenerate neutrino species
corresponds
to allowed BBN models with  a baryon fraction as large
as $\Omega_b h_{50}^2 \la 0.25$ (cf.~Figure \ref{fig:8}). 

\subsection{Neutrino Mass Constraint}

Figure \ref{fig:12}  summarizes the neutrino
contribution to the closure density $\Omega_\nu$ as a function
of neutrino degeneracy and mass.   This figure assumes the plausible
model of nearly degenerate $\nu_\mu$ and $\nu_\tau$ masses and
negligible $\nu_e$ mass.  For this figure $\Omega_\nu$
refers to the combined contributions from both
 $\nu_\mu$ and $\nu_\tau$.
The contribution changes
for large degeneracy due to the lower present-day neutrino
temperature.  This figure can be used
to constrain the masses of the $\nu$ and $\tau$  neutrino types in different
cosmological models.  For example, if we assume a 
model with $\Omega_b h_{50}^2 = 0.1$, $\Omega_\Lambda = 0.6$,
and $\Omega_\nu = 0.3$, then we would find that 
the masses of both the $\nu_{\mu}$ and  $\nu_{\tau}$ must be $\la$2 eV,
 if these two species are 
to provide the neutrino contribution to the closure density.  

Studies of large scale structure also constrain
neutrino masses and degeneracy 
in hot-dark-matter (HDM)  
and mixed-dark-matter (MDM) models.
Indeed, at least some neutrino mass may presently be 
required to account for the observed power spectrum of 
galactic and microwave background structure.  It has been argued 
(\cite{primack}) from considerations of structure formation 
in the early universe that two neutrino  flavors ($\num, \nut$) may have
a rest mass of $2.4$ eV, 
compatible with all neutrino oscillation experiments.  
This postulate solves the main problem of
cold dark matter (CDM) models, i.e.  production of too  much 
structure on small scales.  Furthermore, Larsen \& Madsen (1995)
argue  that neutrino degeneracy is required in 
a MDM model with 2.4 eV neutrinos to obtain an optimum fit to
the power spectrum.. 
If we take 2.4 eV as the given mass
of $\num$ and $\nut$, then  for a $\Omega_\nu \la 0.9$
$\Omega_b h_{50}^2 = 0.1$
MDM cosmology we would deduce from Figure \ref{fig:12}
that the maximum degeneracy  
for two species with this mass would correspond
to $\xinumt \la 2.5$ similar to the values used
in Larsen \& Madsen (1995).  We suggest that in light of the present results
a similar study of the galactic power spectrum for $\xinumt \approx 11$
 and $m_\nu \sim 0.1$ eV should be undertaken as well.

\subsection{Cosmic Microwave Background Constraint}
Perhaps, the most stringent remaining constraint on neutrino degeneracy
comes from its effect on the cosmic microwave background.
Several recent works (\cite{kinney}; \cite{lesgourgues}; 
\cite{hannestad}) have shown that
neutrino degeneracy can dramatically alter the power spectrum of the CMB.
The essence of this constraint is that
degenerate neutrinos increase the energy density in
radiation at the time of photon decoupling in addition to delaying
the time of matter-radiation energy-density equality as discussed above.  
One effect of this is to increase the amplitude of the first
acoustic peak in the CMB power spectrum at $l \approx 200$.
For example, based upon a $\chi^2$ analysis (\cite{lineweaver}) of 19 experimental
points and window functions, Lesgourgues and Pastor (1999)
concluded that $\xi_\nu \le 6$ for a single degenerate neutrino species.
   
However, in the  existing CMB constraint calculations 
(Kinney and Riotto 1999: Lesgourgues and Pastor 1999; Hannestad 2000)
 only small degeneracy parameters with the standard relic neutrino
temperatures were utilized in their derived constraint.
Hence, the possible effects of a diminished relic neutrino temperature 
for large neutrino degeneracy need to be reconsidered.  To investigate this
we have done calculations of the CMB power spectrum, 
$\Delta T^2 = l(l+1)C_l/2 \pi$
based upon the CMBFAST code of Seljak \& Zaldarriago (1996).

For the optimum neutrino-degenerate models ($\xinumt \approx 11$)
and a neutrino contribution $\Omega_\nu \le 0.25$
we deduce from the solid curve on  Figure \ref{fig:9} that 
the neutrino mass is $m_{\nu_{\mu,\tau}} \le 0.1$ eV and therefore
unimportant during the photon decoupling epoch. Therefore,
we only consider massless neutrinos here.
For massless neutrinos it can be proven (\cite{lesgourgues})
that the only effect of neutrino degeneracy on the CMB 
is to increase the background
pressure and energy density of relativistic particles (cf.~Eqs.~1-3).
We have in this way explicitly modified CMBFAST code to account for the 
contribution of massless degenerate neutrinos with a relic temperature
ratio $y_\nu = T_\gamma/T_\nu$ as given in Figure \ref{fig:4}.

We have evaluated $\chi^2$ for fits to 
the CMB power spectrum, based upon
the "radical compression" technique as described in Bond, Jaffe \& Knox (2000).
We have used the latest 69 observational points and window functions 
available from the web page given in that paper.  
The advantage of this approach is that
the non-Gaussian experimental uncertainties 
in the power spectrum are correctly weighted in the evaluation of the 
goodness of fit.

For the purposes of the present study, we take as a benchmark the "All" case
best fit $\Omega = 1$ model of Dodelson \& Knox (2000) who derived cosmological
parameters based upon this same data set and compression technique.  
Although there is some
degeneracy in the cosmological parameters they deduced  an optimum 
fit to the power spectrum for
 $\Omega_b h^2 = 0.019$, $H_0 = 65$ km sec$^{-1}$ Mpc$^{-1}$, 
$\Omega_\Lambda = 0.69$, $\Omega_M = 0.31$,
 $\tau = 0.17$, and  $n = 1.12$,
where $\Omega_M$ is the total matter contribution,
$\tau$ is the reionization parameter, and $n$ is the "tilt" of the 
power spectrum.
This benchmark is plotted as the dashed curve in Figure \ref{fig:14}.
For this case we find $\chi^2 = 101$.  [Note that our $\chi^2$
is slightly different from that quoted in Dodelson \& Knox (2000) because
we use different binning of the power spectrum]. 
For comparison the  "radical
compression" of the 
CMB data into 14 bins used in this work is also shown (\cite{bond}).

Rather than to do an exhaustive parameter search we have taken an 
approach similar to Lesgourgues \& Pastor (1999).  That is, we fix several 
representative cosmological
models and then study their goodness of fit to the
CMB data.  The best
case for large neutrino degeneracy will be for a value of the degeneracy
parameter $\xinumt$ such that neutrino decoupling occurs just before the
QCD phase transition.  This is the value for which the relic neutrino energy density
is a local minimum (cf. Figure \ref{fig:13}).
 For the present work this corresponds to 
 $\xinumt = 11.4$, $\xinue = 0.73$, $\omegab \h502 = 0.187$ models. 
In what follows we fix $\xinumt$, $\xinue$, and $\omegab \h502$ at these values
and refer to this as the large degeneracy model.

We have found [as did Lesgourgues \& Pastor (1999)] that 
the currently favored $\Omega_\Lambda = 0.7$
models give a poor fit to the data even with no degeneracy.
Adding neutrino degeneracy to an $\Omega_\Lambda = 0.7$ model
only makes the fit worse. 
The main problem is that
 the first acoustic peak increases in amplitude 
and moves to larger $l$.  Hence, even though a local minimum develops
for large degeneracy, the $\chi^2$ is substantially increased and
large neutrino degeneracy is probably ruled out for 
$\Omega_\Lambda = 0.7$ models.

For smaller $\Omega_\Lambda$ a local minimum  develops in the
$\chi^2$ for both small values of degeneracy
$\xinumt \approx 1$ and large degeneracy $\xinumt = 11.4$. 
As pointed out in Lesgourgues \& Pastor (1999), the best
case for neutrino degeneracy is with $\Omega_\Lambda = 0$ models.
However, those models are probably ruled out by observations  of
type Ia supernovae at high redshift (\cite{garnavich};
 \cite{perlmuttera};
 \cite{perlmutterb};
\cite{riess}).
At the $3 \sigma$ confidence level for $\Omega = 1$ models, 
the type Ia Supernova data are consistent with $\Omega_\Lambda = 0.7 \pm 0.3$
Hence, we take $\Omega_\Lambda = 0.4$ as a plausible 
cosmological model which is marginally consistent with the type Ia results. 
Nevertheless, for purposes of  illustration, we have also
made a search for
optimum parameters for matter dominated $\Omega_\Lambda = 0$, 
$\Omega = 1$ models.

The reason low $\Omega_\Lambda$ models are favored is that they shift
the first acoustic peak back to lower $l$.
Larger values of $H_0$ also slightly improve the
fit by shifting the first acoustic peak to lower $l$ and decreasing
the baryon density for fixed $\omegab \h502$ which lowers the
peak amplitude.  We take $H_0 = 65 \pm 10$ ($h_{50} = 1.3 \pm 0.2$)
as a reasonable range 
(\cite{dodelson}), and therefore utilize
$h_{50} = 1.5$ as the optimum Hubble parameter for the neutrino-degenerate models. 
This implies that $\Omega_b = 0.084$ for the large degeneracy models. 
The ionization parameter does not particularly
 help the fits as it mainly serves to decrease the amplitude of both
the first and second peaks in the power spectrum.
We therefore  set $\tau = 0$ for the large degeneracy models.
The only  remaining adjustable parameter of the fits is then the tilt parameter $n$. 
Values of $n$ slightly below unity also help with the amplitude and
 location of the first acoustic peak. 

The solid line on Figure \ref{fig:14} shows  a  $\Omega_\Lambda = 0.4$
model for which $n = 0.78$.  For this fit $\Delta \chi^2 = 27$  which makes
this large degeneracy model marginally consistent with the data at a 
level of $5.2 \sigma$.
The dotted line in Figure \ref{fig:14} shows the matter dominated 
$\Omega_\Lambda = 0$ best fit model with $n = 0.83$.
For this fit $\Delta \chi^2 = 9$  which makes
this large degeneracy model consistent with the data at 
the level of $3 \sigma$.

As can be seen from Figure \ref{fig:14}, a model with large neutrino
degeneracy seems marginally acceptable based upon the presently uncertain
power spectrum.  
The main differences in the fits between the large degeneracy models
and our adopted benchmark model are that the first
peak is shifted to slightly higher $l$ values
and the second peak is somewhat suppressed.  It thus becomes
important to quantify the amplitude of the second peak in order
to constrain the large degeneracy models proposed herein.

Indeed, after the present fits were completed a
suppression of the second acoustic peak in the power  spectrum was  
reported  in the 
 high-resolution BOOMERANG (\cite{boomerang}; \cite{lange})
and MAXIMA-1 (\cite{hanany}: \cite{balbi}) results.  
 We have not yet analyzed the goodness of fit
to these data as the experimental window functions are not yet available.  
In a subsequent paper we will examine the implications of those data in detail.

For purposes of illustration, however,
 we compare the fit models of Figure \ref{fig:14} with
the published BOOMERANG and MAXIMA-1 power spectra
in Figure \ref{fig:15}.
Here one can clearly see that the suppression of the second
acoustic peak in the observed power spectrum is consistent
with our derived neutrino-degenerate models.
In particular, the MAXIMA-1 results are in very good agreement with the 
predictions of the neutrino-degenerate cosmological
models described herein.  
There is, however, a calibration uncertainty between these
two sets (\cite{hanany}).  If one only considers
the BOOMERANG results alone, the diminished amplitude of the first
acoustic  peak probably tightens the constraint for low
neutrino degeneracy models (cf.~\cite{hannestad})
although even for this set alone,  a high degeneracy
model is probably still acceptable (\cite{lesgourgues00}). 
It is clear, that these new data sets will substantially improve the
goodness of fit for the neutrino-degenerate models (\cite{lesgourgues00}).  
Moreover, both data sets seem to require an increase in
the baryonic contribution to the closure density
 as allowed in our neutrino-degenerate models.

\section{CONCLUSIONS}

  We have discussed how the relic neutrino temperature is substantially 
diminished in cosmological models with a large neutrino degeneracy.
We have shown that all of the BBN light-element abundance constraints
(assuming some destruction of $^7$Li) can be satisfied for 
a single-parameter family of cosmological models in which significant neutrino
degeneracies and large values of $\Omega_b h^2$ can exist.
The requirement that large scale structure become nonlinear
in sufficient time can also be satisfied for models
with either moderate degeneracy 
($\xinumt \la 6.2$ and $\omegab \h502 \la  0.22$) or  
large neutrino degeneracy
($11.4 \la \xinumt \la 13$ and  $0.187 \la \omegab \h502 \la  0.25$).
We have also shown that even the constraint from 
neutrino-degeneracy effects on fluctuations
of the cosmic microwave background temperature
may be marginally avoided for models with
$\Omega_\Lambda \la 0.4$,  $\xinumt \approx  11$, and 
$\omegab \h502 \approx 0.2$.

At present, the power spectrum of the CMB is the most
stringent constraint.  Nevertheless,
neutrino-degenerate models can be found which are 
marginally consistent at the 3-5$\sigma$ level.
This tight constraint is due, at least in part,
to a suppression of the second acoustic peak in the spectrum.
It is therefore encouraging that 
the recent BOOMERANG and MAXIMA-1 results suggest that such
a suppression in the second acoustic peak may indeed
occur in agreement with the expectations 
of the large neutrino degeneracy, high $\omegab$ 
 models proposed here.
   
Thus, high resolution microwave background observations
become even more important as a means to quantify the limits
to (or existence of) possible cosmological neutrino degeneracy.  
Based upon the current analysis, we conclude that
all of the cosmological constraints on large neutrino degeneracy
 are marginally satisfied when a
careful accounting of the neutrino decoupling and relic neutrino temperature
is made.  It will, therefore, be
most interesting to see what further constraints can be placed
on this possibility from the soon to be available
 space-based high resolution CMB observations such as
the NASA MAP and ESA Planck missions.

\acknowledgments
One of the authors (GJM) wishes to acknowledge the hospitality
of the National Astronomical Observatory of Japan where much of this
work was done.
This work has been supported in part by the Grant-in-Aid for
Scientific Research (10640236, 10044103, 11127220, 12047233) 
of the Ministry of Education, Science, Sports, and Culture of Japan, 
and also in part by NSF Grant (PHY-9901241 at OSU) along with
DoE Nuclear Theory Grant (DE-FG02-95-ER40394 at UND).

%
%
%
%

%
%
{
 \setcounter{enumi}{\value{figure}}
 \addtocounter{enumi}{1}
 \setcounter{figure}{0}
 \renewcommand{\thefigure}{\theenumi(\alph{figure})}
\begin{figure}
\plotone{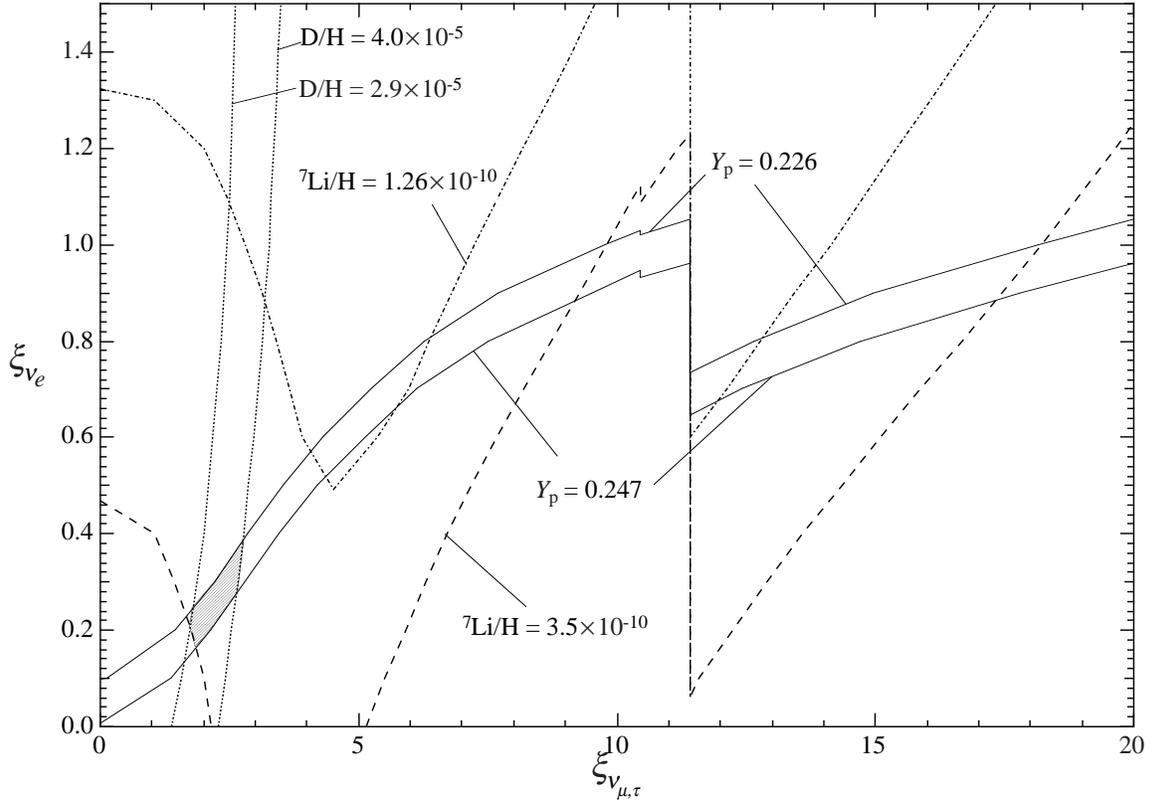}
\caption{Contours of allowed values in the $\xinue~ - ~\xinumt$ plane
for $\omegab \h502$ = 0.1, based upon the various light-element abundance
constraints as indicated.  The hatched region depicts the allowed parameters
consistent with all light element constraints for this value of $\omegab \h502$.}
\protect\label{fig:7a}
\end{figure}
\begin{figure}
\plotone{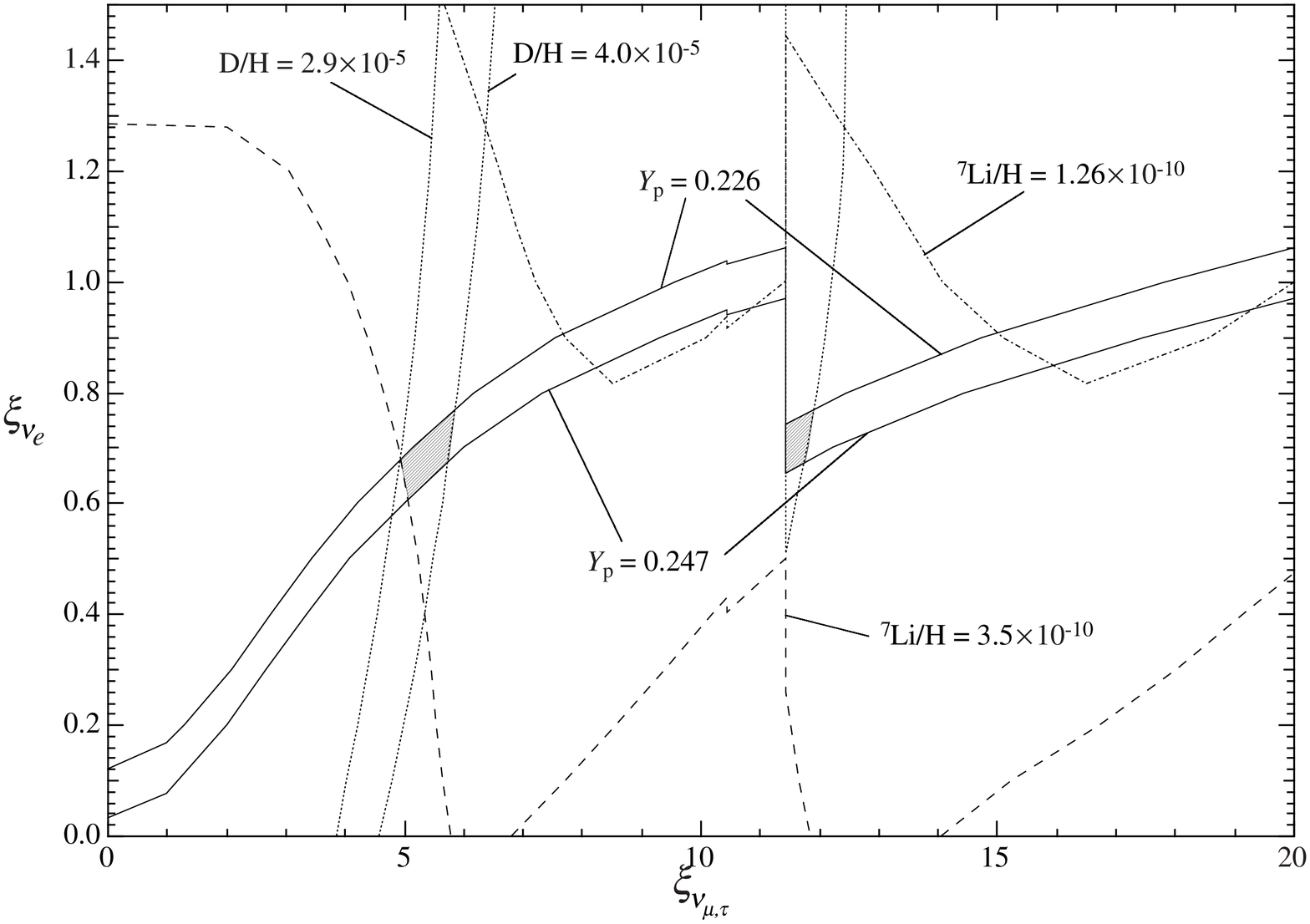}
\caption{Same as Fig7a, but for $\omegab \h502$ = 0.2.}
       \protect\label{fig:7b}
\end{figure}
\begin{figure}
\plotone{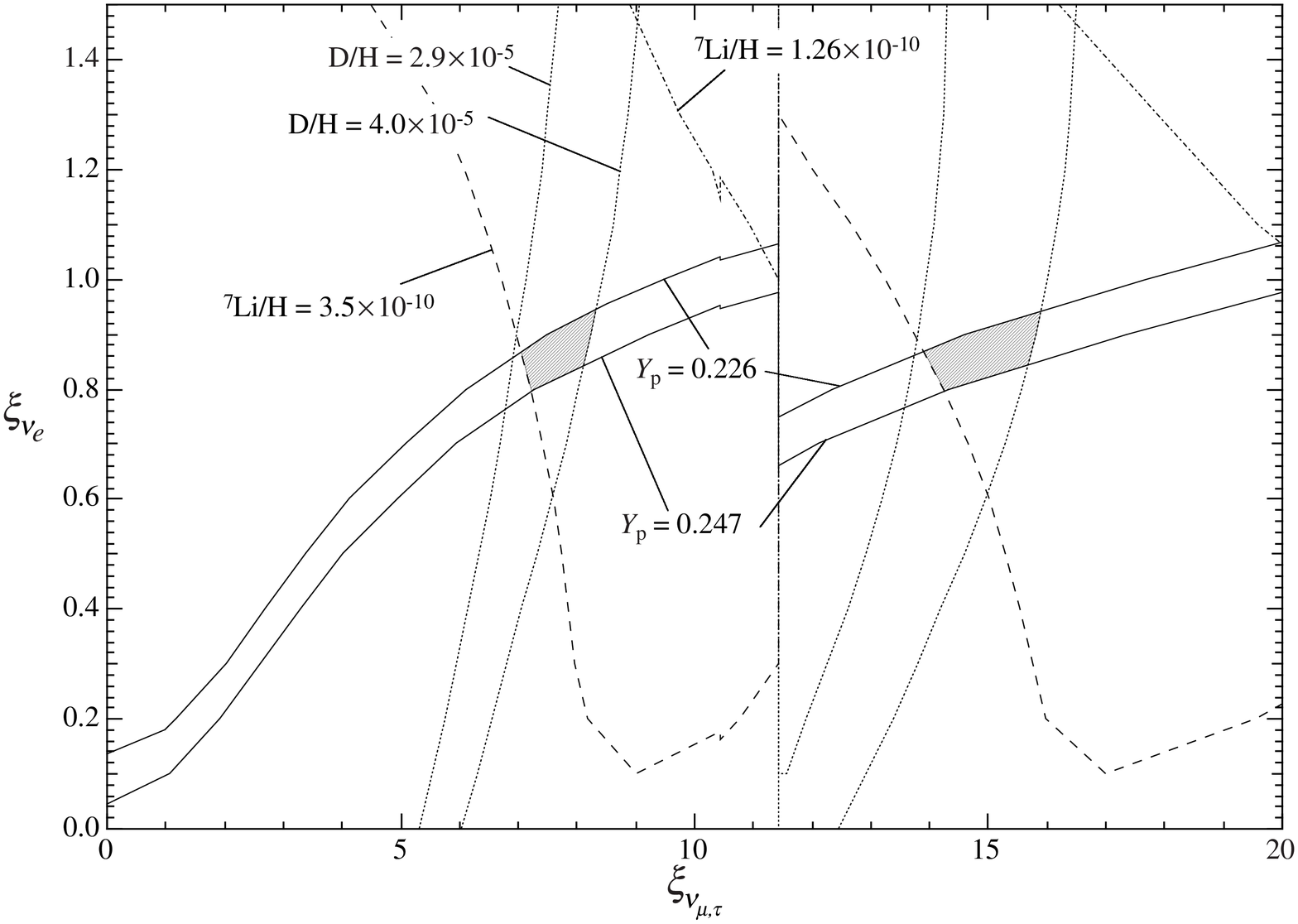}
\caption{Same as Fig7a, but for $\omegab \h502$ = 0.3.}
        \protect\label{fig:7c}
\end{figure}
\setcounter{figure}{\value{enumi}}
 }
\begin{figure}
\plotone{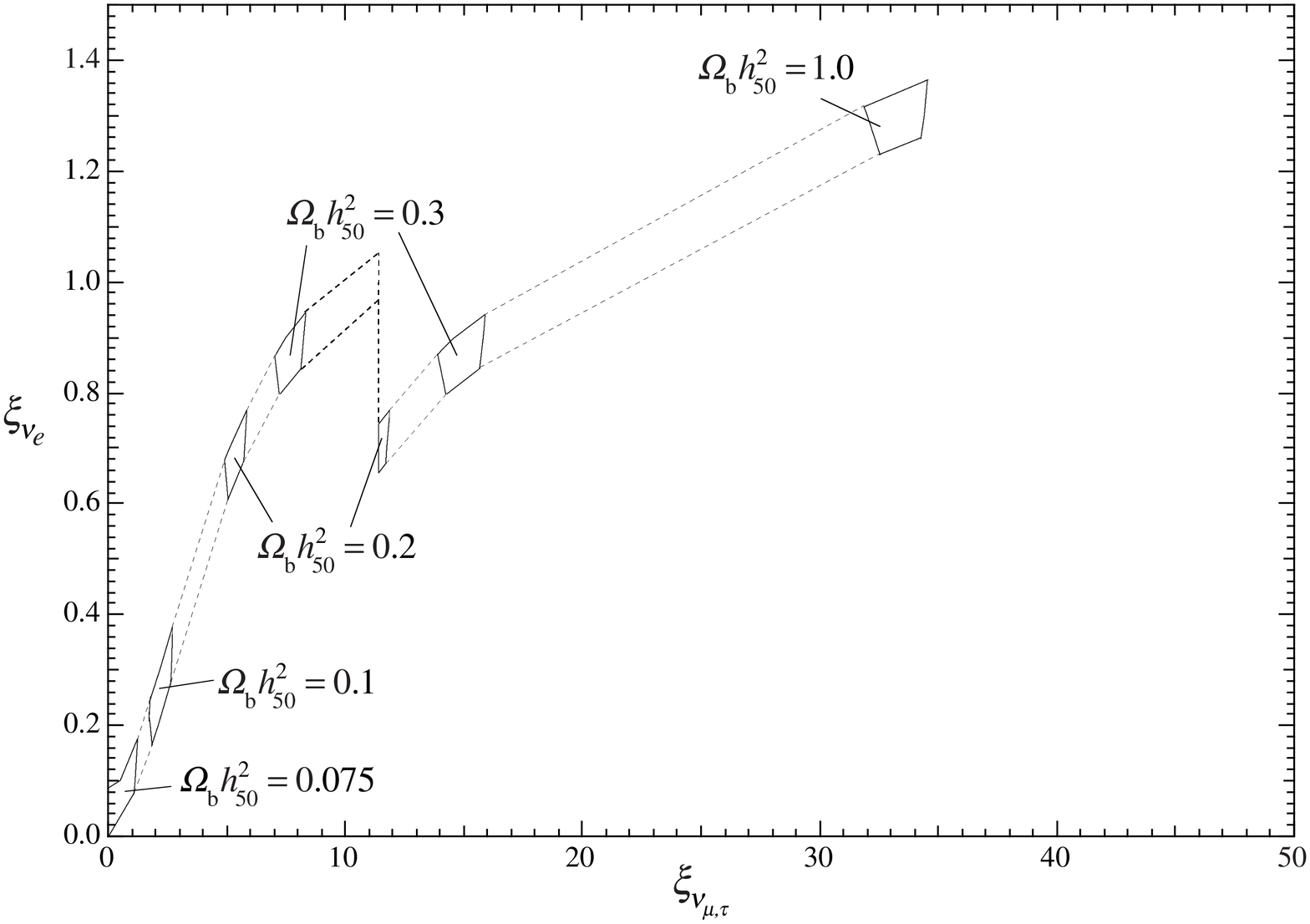}
\caption{Allowed values of $\xinue$ and $\xinumt$ for which the
constraints from light element abundances are satisfied for
values of  $\omegab
\h502 =$ 0.075, 0.1, 0.2, 0.3 and 1.0 as indicated. 
For large values of $\omegab \h502 > 0.3$
the only allowed regions are 
 for the large values $\xinumt > 20$. }
\protect\label{fig:8}
\end{figure}
\vfill\eject
\begin{figure}
\plotone{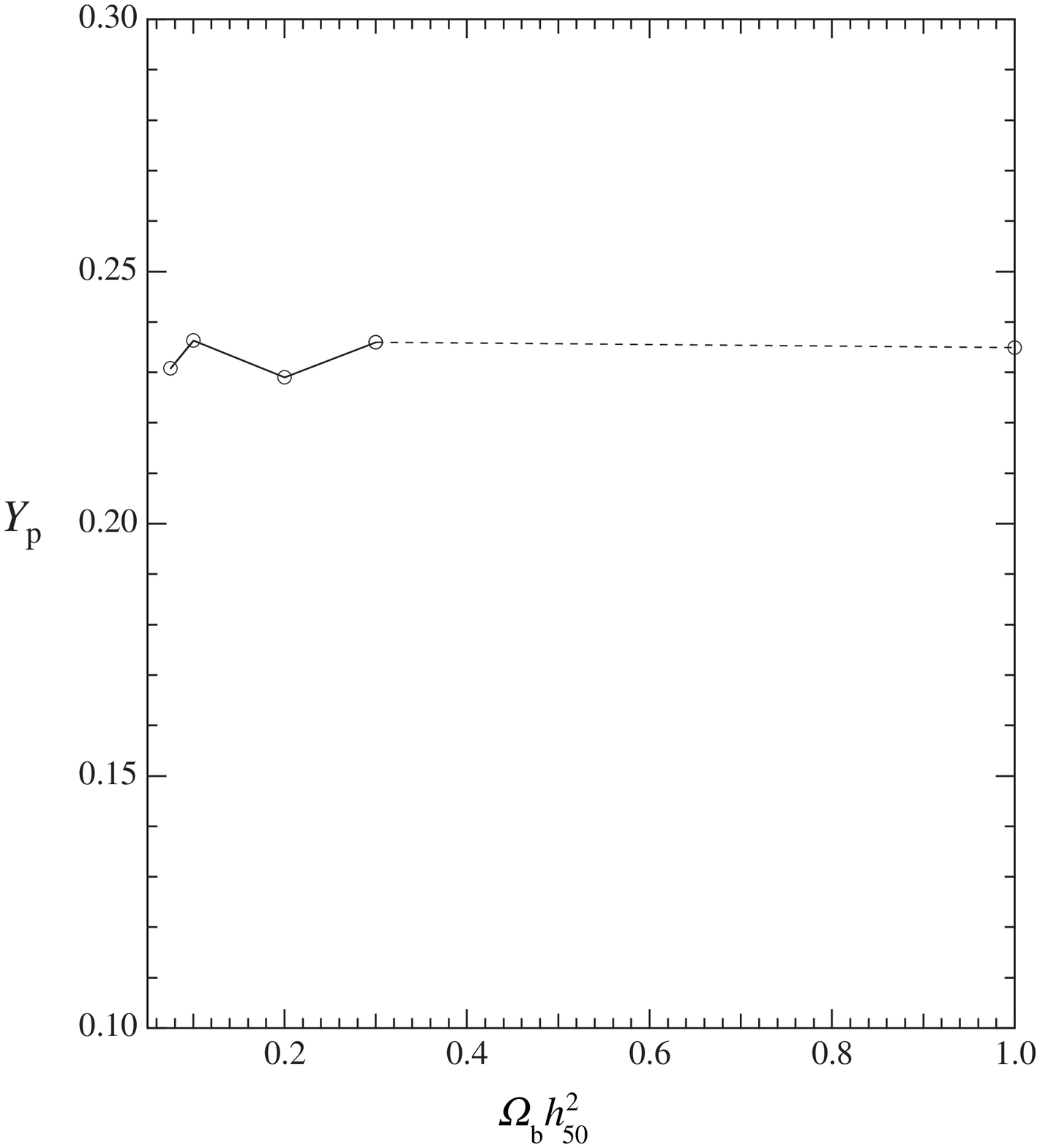}
\caption{The predicted Helium abundance for allowed 
neutrino-degenerate models
as a function of $\omegab \h502$. The values
of $\xinue$ and $\xinumt$ were taken from the central value in
the allowed region determined by $\omegab \h502$ in 
Fig.~\protect\ref{fig:8}. Note that \yp is sensitive to the choice of 
$\xinue$ in the allowed region.}
\protect\label{fig:9}
\end{figure}
\begin{figure}
\plotone{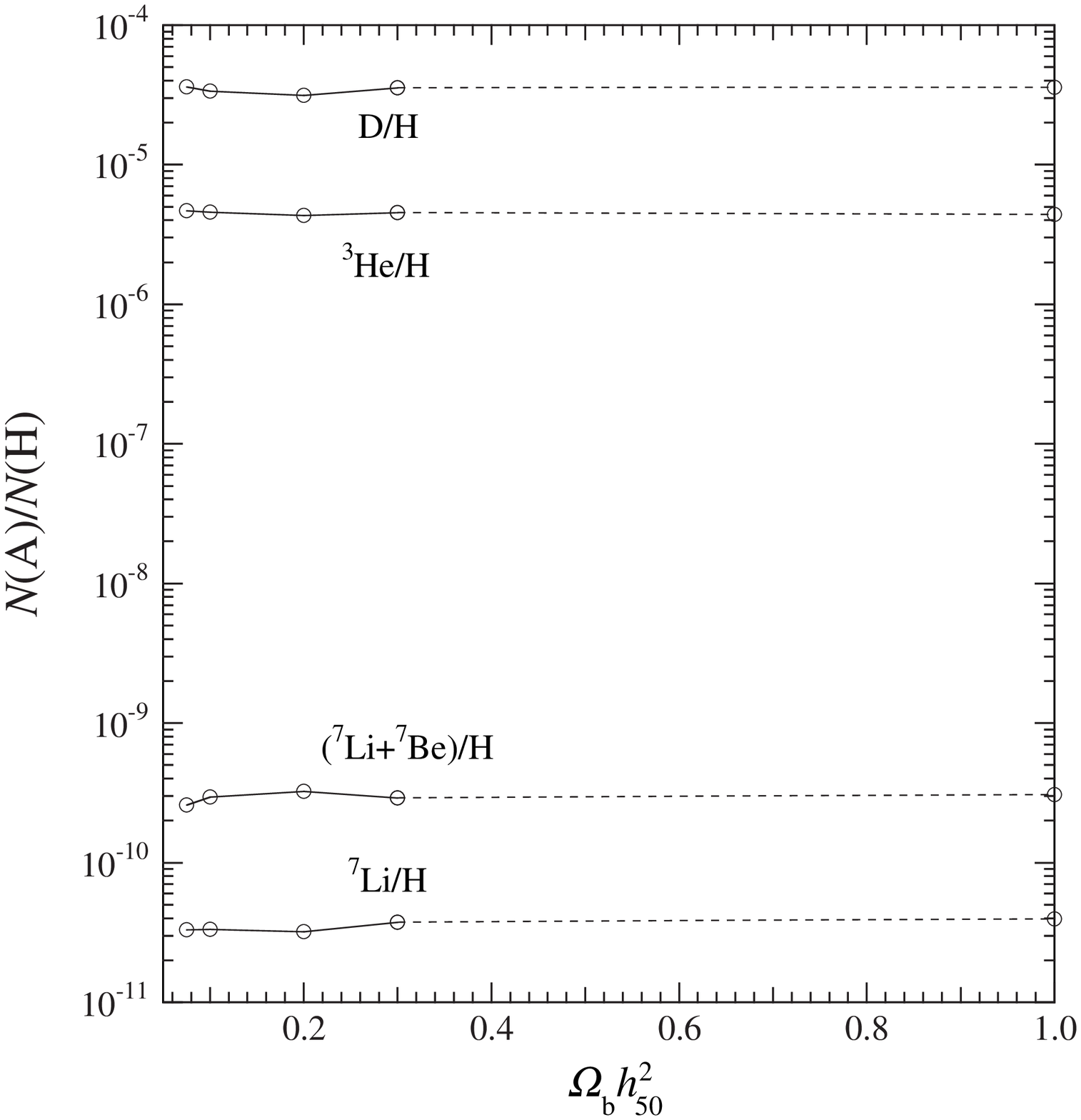}
\caption{The predicted D/H, $^{3}$He/H, and
 total $A=7$ and $^7$Li abundances for allowed 
neutrino-degenerate models
as a function of $\omegab \h502$. The values
of $\xinue$ and $\xinumt$ were taken from the central value in
the allowed region determined by $\omegab \h502$ in Fig.~\protect\ref{fig:8}.}
\protect\label{fig:10}
\end{figure}
\begin{figure}
\plotone{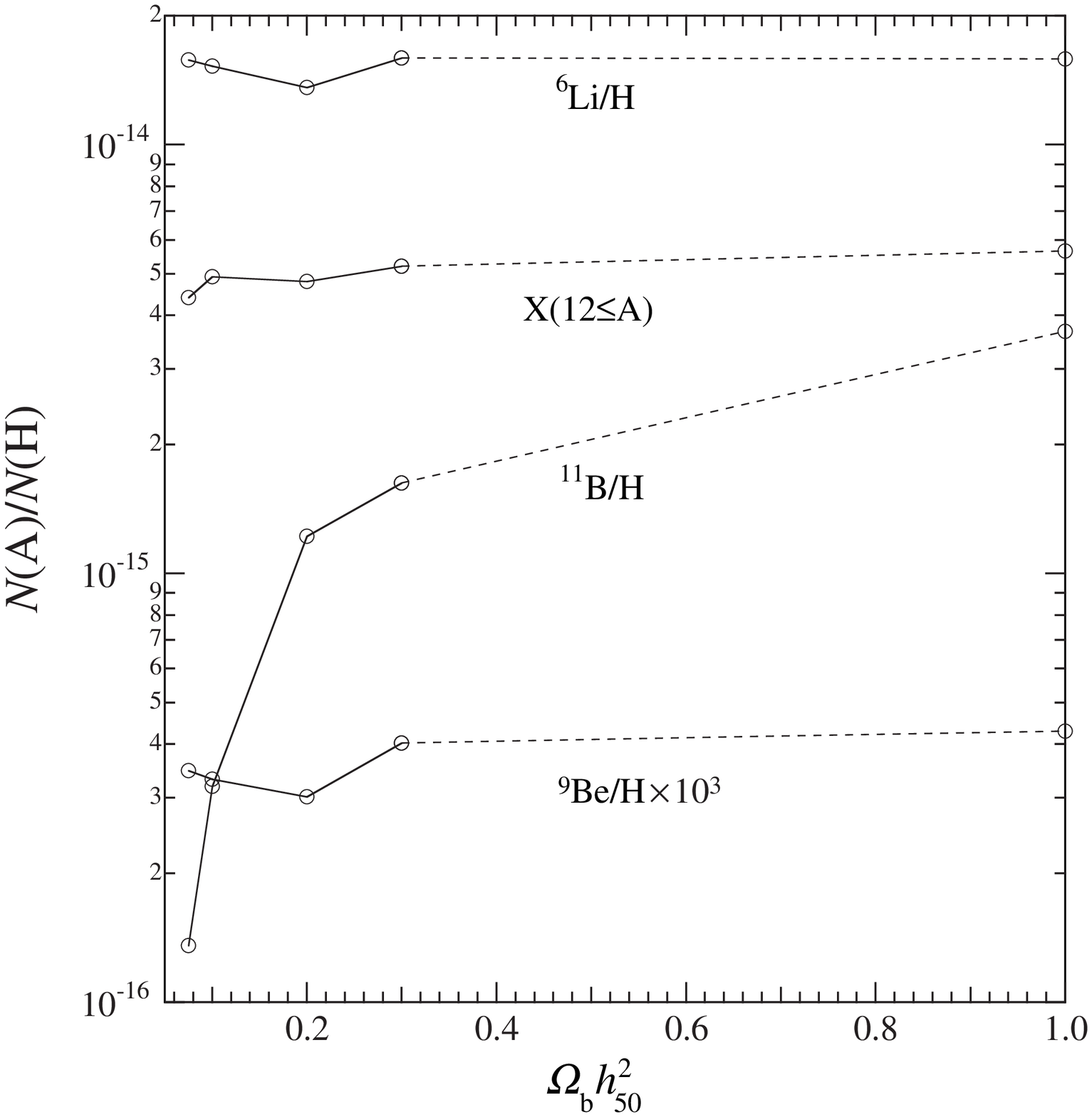}
\caption{The predicted $^6$Li, $^9$Be, $^{11}$B and $A \ge 12$  heavier
element abundances as a function of $\omegab \h502$. 
The values of $\xinue$ and $\xinumt$ were taken from the central value in
the allowed region determined by $\omegab \h502$ in Fig.~\protect\ref{fig:8}.}
\protect\label{fig:11}
\end{figure}
\begin{figure}
\epsscale{0.9}
\plotone{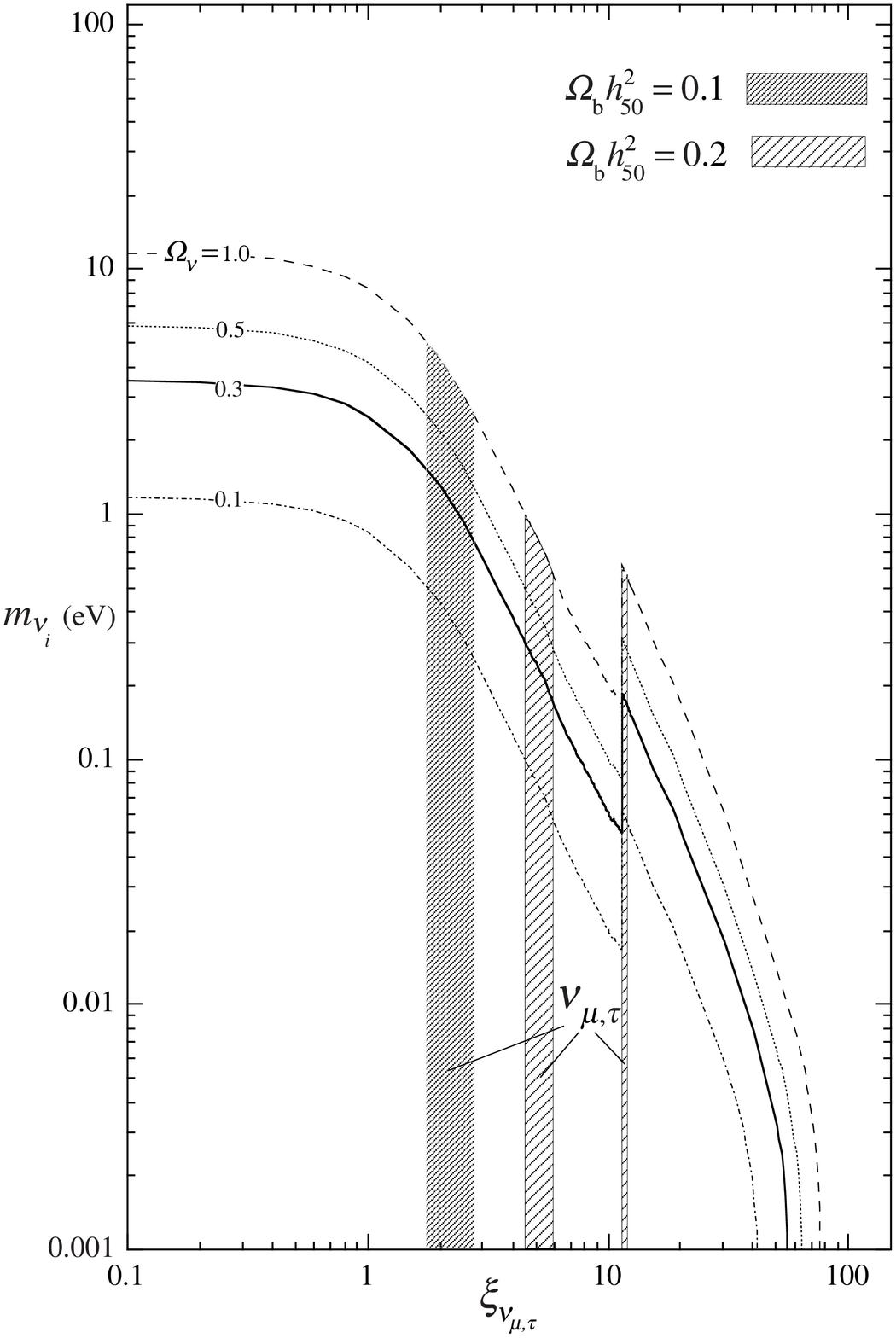}
\caption{Contours of equal present energy density of massive
degenerate neutrinos as a function of the degeneracy $\xinumt$ and
neutrino mass $m_\nu$ for $\Omega_\nu$ =  0.1, 0.3, 0.5, and
1.0 as indicated. Each curve corresponds to different value of
$\Omega_\nu$ as indicated. Shaded regions depict the allowed range of
degeneracy for the two indicated values of  $\omegab \h502 = 0.1$, 
and $0.2$ as shown in Fig.~8.}
\protect\label{fig:12}
\end{figure}
\begin{figure}
\plotone{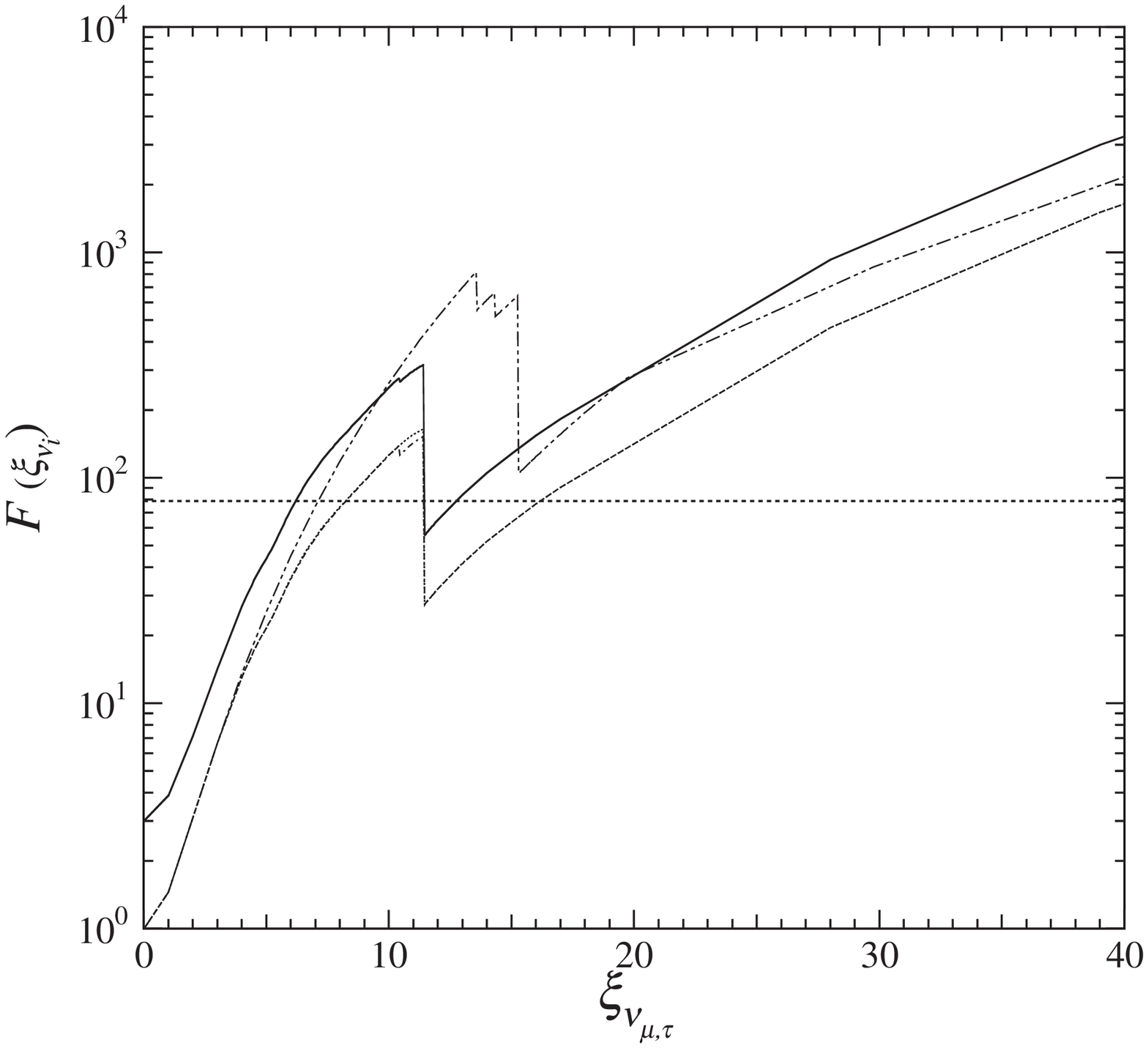}
\caption{Calculated neutrino energy density factors $F(\xi_{\nu_i})$ as
a function of degeneracy parameter for the three neutrino species.
The dotted and dot-dashed curves display respectively $F(\xi_{\nu_\mu})$ 
and $F(\xi_{\nu_\tau})$ for the cases 
in which only one neutrino species is degenerate.
Since our interesting parameter regions in Figure \ref{fig:8} satisfy 
$\xinue \ll \xinumt$, only $\nu_{\mu,\tau}$ contribute significantly 
to the total $F(\xi_\nu)$ (solid curve).
The dashed horizontal line indicates the constraint on neutrino degeneracy
from the requirement that sufficient structure develops by the present 
time. We also show the previous estimate (two-dot dashed curve) of 
Kang \& Steigman (1992).}
\protect \label{fig:13}
\end{figure}
\begin{figure}
\plotone{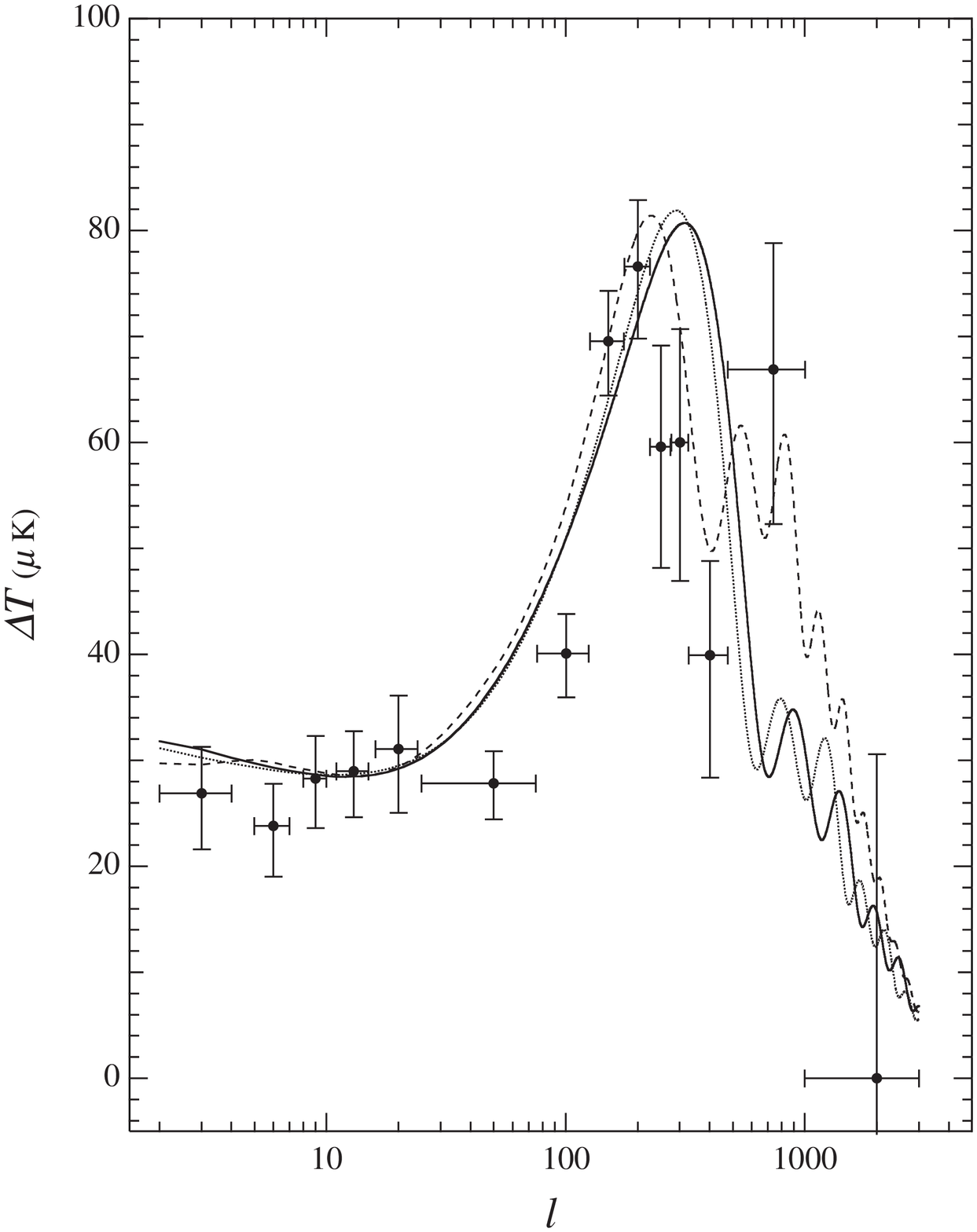}
\caption{ CMB power spectrum compared with calculated $\Omega = 1$ models.  
The points show the binning of 69 experimental
measurements based upon the radical compression method of Bond et al. (2000).  
The dashed line shows the optimum model of Dodelson \& Knox (2000).  
The solid line shows an $\Omega_\Lambda = 0.4$ model with three degenerate 
neutrinos $\xinumt = 11.4$, $\xinue = 0.73$  and $\omegab \h502 = 0.187$
as described in the text.
The dotted line is for an $\Omega_\Lambda = 0$ model with
the same degeneracy  and baryonic parameters.}
\protect \label{fig:14}
\end{figure}
\begin{figure}
\plotone{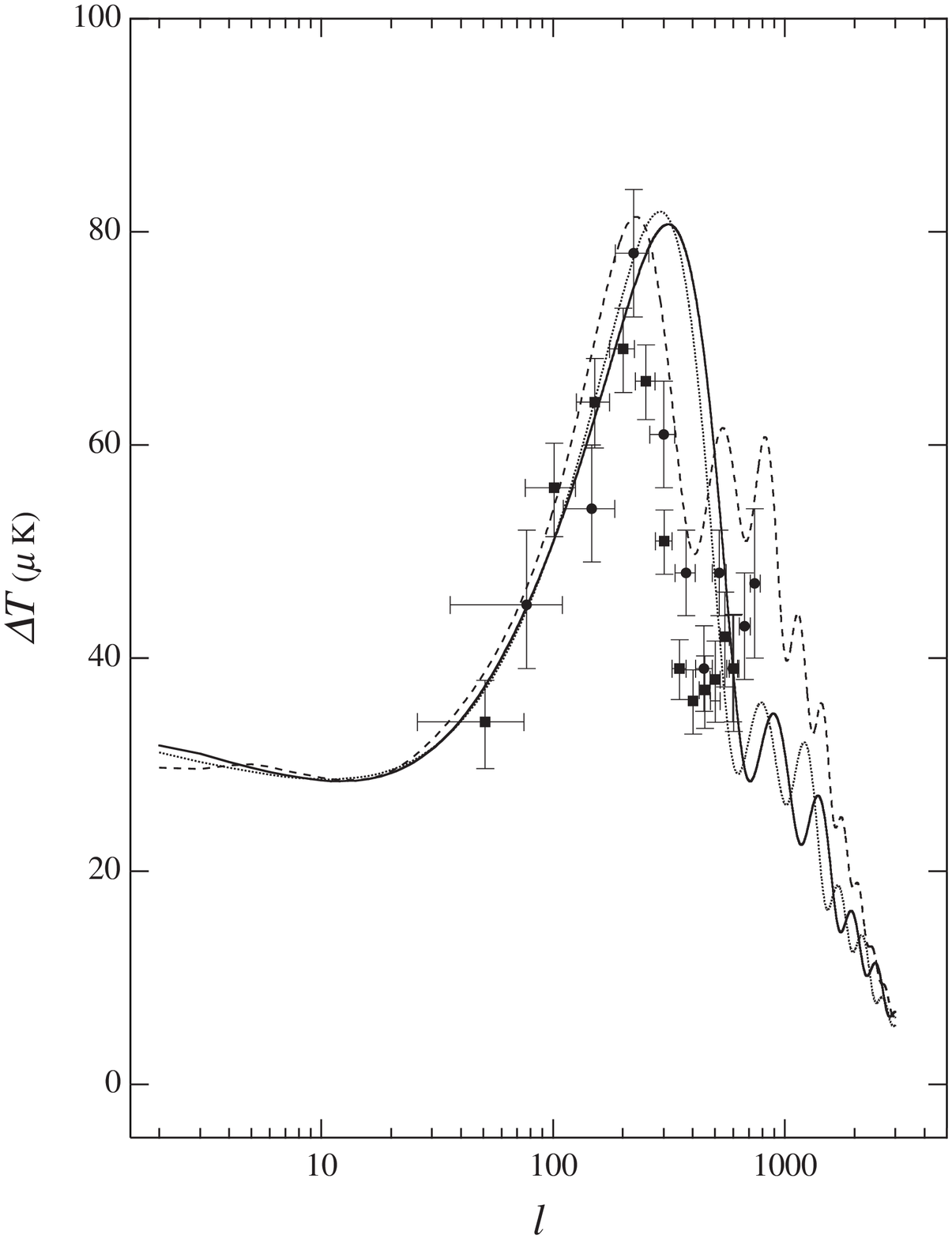}
\caption{ CMB power spectrum  from the recent 
MAXIMA-1 (circles) and BOOMERANG  (squares) 
binned data compared with calculated $\Omega = 1$ models
of Figure 14.  Note that the suppression of the second acoustic
peak in the data is consistent with that predicted by the 
neutrino-degenerate models.}
\protect \label{fig:15}
\end{figure}

\end{document}